\documentclass[12pt]{article}
\pdfoutput=1
\usepackage{putex}
\usepackage{graphicx}
\usepackage{epstopdf}
\usepackage{enumerate}
\usepackage[nosort]{cite}
\usepackage{tensor}
\usepackage{slashed}
\usepackage{feynmf}
\usepackage{hyperref}
\usepackage{float}
\usepackage[bottom]{footmisc}
\usepackage[utf8]{inputenc}
\usepackage{amsmath}
\usepackage[usenames,dvipsnames,svgnames,table]{xcolor}
\usepackage[bf,hang,footnotesize]{caption}
\usepackage{bbold}

\numberwithin{equation}{section}
\hypersetup{
	pdfstartview={FitH},    % fits the width of the page to the window
	pdftitle={S-Matrix Path Integral Approach to Symmetries and Soft Theorems},    % title
	pdfauthor={Kim, Kraus, Monten, Myers},     % authors
	colorlinks=true,       % false: boxed links; true: colored links
	linkcolor=blue,          % color of internal links
	citecolor=blue!59!black! ,        % color of links to bibliography
	filecolor=magenta,      % color of file links
	urlcolor=blue!75!black,           % color of external links
	linktoc=all
}

\makeatletter
\g@addto@macro\bfseries{\boldmath}
\makeatother

\numberwithin{equation}{section}

\newcommand{\eq}[2]{\begin{align}\label{#1}#2\end{align}}

\def\wb{\overline{w}}

\newcommand {\be} {\begin {equation}}
\newcommand {\ee} {\end {equation}}

\newcommand{\p}{\partial}

\newcommand\rt{{\rightarrow}}
\def\eps{\epsilon}

\newcommand{\rf}[1]{(\ref{#1})}
\newcommand{\rff}[1]{\ref{#1}}

\newcommand{\zb}{\overline{z}}

\definecolor{greenC}{rgb}{0.0, 0.38, 0.18}

\newcommand{\phib}{\overline{\phi}}

\newcommand{\Ab}{\overline{A}}

\newcommand{\xh}{\hat{x}}

\newcommand{\dr}{d}

\newcommand{\xv}{\vec{x}}
\newcommand{\scrI}{\mathcal I}
\newcommand{\pv}{\vec{p}}
\newcommand{\normord}[1]{\mathopen{:}#1\mathclose{:}}

\begin{document}

\institution{UCLA}{ \quad\quad\quad\quad\quad\quad\quad\ ~ \, $^{1}$Mani L. Bhaumik Institute for Theoretical Physics
		\cr Department of Physics \& Astronomy,\,University of California,\,Los Angeles,\,CA\,90095,\,USA}

\title{ S-Matrix Path Integral Approach to Symmetries and Soft Theorems}

\authors{Seolhwa Kim$^{1}$, Per Kraus$^{1}$, Ruben Monten$^{1}$, Richard M. Myers$^{1}$}
	
\abstract{  We explore a formulation of the S-matrix in terms of the path integral with specified asymptotic data, as originally proposed by Arefeva, Faddeev, and Slavnov. In the  tree approximation the  S-matrix is equal to the exponential of the classical action evaluated on-shell.   This formulation is well-suited to questions involving asymptotic symmetries, as it avoids reference to non-gauge/diffeomorphism invariant bulk correlators or sources at intermediate stages. We show that the soft photon theorem, originally derived by Weinberg and more recently connected to asymptotic symmetries by Strominger and collaborators, follows rather simply from invariance of the action under large gauge transformations applied to the asymptotic data.  We also show that this formalism allows for efficient computation of the S-matrix in curved spacetime, including particle production due to a time dependent metric. 
 }
	
	\date{}
	
	\maketitle
	\setcounter{tocdepth}{2}
	\begingroup
	\hypersetup{linkcolor=black}
	\tableofcontents
	\endgroup
	
%%%%%%%%%%%%%%%%%%%%%%%%%%%%%%%%%%%%%%%%%%%%%%%%%%%

\section{Introduction}

Given the fundamental role of the S-matrix in connecting quantum field theory to experiment, it is valuable to explore and develop its alternative formulations.  The most popular formulation is  the LSZ prescription, in which S-matrix elements are extracted from the poles of off-shell correlators of quantum fields.   Among other virtues, one powerful property of LSZ is that it applies equally well to the scattering of elementary particles and bound states.  To find an alternative prescription, it is useful to make an analogy with the computation of boundary correlators in the AdS/CFT correspondence. Working in the bulk, there are two well known prescriptions.   In the GKP/W prescription \cite{Gubser:1998bc,Witten:1998qj} we are instructed to compute the supergravity/string theory path integral as a functional of asymptotic boundary conditions near the timelike boundary of AdS.   These boundary data, $J(x)$, are then identified with sources in the dual CFT.  For a CFT operator ${\cal O}$ of scaling dimension $\Delta$ the corresponding bulk field behaves as $\phi_\text{bulk} \sim r^{\Delta-d} J$ in AdS$_{d+1}$.   This leads to the  basic statement of the AdS/CFT correspondence

\eq{i1}{  Z_{AdS}[J] &= \int_{\phi_\text{bulk}\sim r^{\Delta - d}J} \! {\cal D} \phi_\text{bulk} \, e^{-I_\text{bulk}[\phi_\text{bulk}]} 
\cr
 =  Z_{CFT}[J] &=\int \! {\cal D} \phi_\text{CFT} \, e^{-I_\text{CFT}[\phi_\text{CFT}] + \int J(x) {\cal O}_\text{CFT}(x) \dr^d x} ~. }
Thus $Z[J]$ serves as a generating functional for boundary correlators,
\eq{i1a}{\langle {\cal O}(x_1) \ldots   {\cal O}(x_n) \rangle_\text{CFT} =  \left[ {\delta \over \delta J(x_1)} \ldots {\delta \over \delta J(x_n)} Z[J] \right]_{J=0}.}

The second prescription, of BDHM  \cite{Banks:1998dd}, involves computing a bulk correlator and then taking the bulk points to the boundary by a suitable limiting process,
\eq{i2}{  \langle {\cal O}(x_1) \ldots   {\cal O}(x_n) \rangle_{\text{CFT}} = \lim_{r_1 , \ldots r_n \rt \infty}  r^{\Delta}_1 \ldots  r_n^{\Delta}  \langle \phi(r_1,x_1) \ldots \phi(r_n,x_n) \rangle_{\text{bulk}}~.}
Since the starting point is a bulk QFT correlator, the formulation \rf{i2} makes it clear that boundary correlators may be computed perturbatively in terms of the usual Feynman diagram expansion, here referred to as Witten diagrams, with the external  lines replaced by bulk-boundary propagators.   On the other hand, the version \rf{i1} is more holographic in spirit in the sense that it only makes reference to boundary data.   For our purposes, another feature of \rf{i1} is that it is well adapted to incorporating asymptotic symmetries, since these are defined to act on asymptotic data.   The two formulations are of course equivalent, as we verify in appendix \ref{sec:AdSCFT} using the same argument that we apply to the S-matrix. 

Coming back to the Minkowski space S-matrix, it is apparent that version \rf{i2} of the AdS/CFT dictionary is analogous to LSZ.   On the other hand, the analog of \rf{i1} was written down long ago by Arefeva, Faddeev, and Slavnov (AFS)\footnote{For a more expansive discussion we recommend Faddeev's chapter in \cite{Balian:1976vq}.   See also \cite{ Faddeev:1980be, Deligne:1999qp, Shrauner:1977sk,Jevicki:1987ax,Adamo:2018srx} for further discussion, and \cite{Adamo:2017nia,Adamo:2021rfq,Gonzo:2022tjm} for recent applications of this or related methods to the study of tree level scattering on curved backgrounds.   }  but remains relatively obscure.     Our objective here is to further develop some aspects of this formalism, with an eye towards application to asymptotic symmetries and related topics.

The idea is to consider the classical action, or more generally the path integral, as a functional of asymptotic boundary data in the far past and future\footnote{More specifically, at timelike or null infinity, depending on whether we are considering massive or massless particles.}.   For the S-matrix, we are interested in field configurations that approach a superposition of plane waves in the far past and future, representing the incoming and outgoing particles.  The boundary conditions we must impose  then consist of fixing the positive(negative) frequency parts of such field configurations in the far past(future).  We denote these parts as $\phib_+$, and $\phib_-$ respectively, so that our boundary conditions on the field $\phi(x)$ are 
\eq{i3}{ \phi_-(x) &\sim \phib_-(x)~,\quad t\rt +\infty \cr
\phi_+(x) &\sim \phib_+(x)~,\quad t\rt -\infty~. }
The observation of \cite{Arefeva:1974jv} is that the path integral, or at tree level the exponential of the on-shell classical action, with these boundary conditions computes the matrix elements of the S-matrix in a basis of suitably normalized coherent states $| \phib_+ \rangle$, defined below,
\eq{i4}{ S[\phib]\equiv \langle \phib_-| \hat{S} | \phib_+ \rangle = \int\! {\cal D} \phi \, e^{i I[\phi,\phib]}~.}
The object $S[\phib] $ serves as a generating functional for S-matrix elements in the usual Fock space basis, in analogy with \rf{i1a}.  Alternatively, an expression for the S-matrix operator in terms of Fock space creation and annihilation operators is given by   
\eq{i5}{ \hat{S} =  \normord{ e^{-iI_\text{bndy}[\hat{\phib},\hat{\phib}]} S[\hat{\phib}] }   }
where $\phib$ is now interpreted as the quantum field $\hat{\phib}$ with its usual free field  mode expansion, and $I_\text{bndy}$ is a boundary contribution to the action that we will discuss in detail later.

In AdS/CFT, boundary terms in the action play an important role in establishing the precise connection between the bulk and boundary partition functions, and the same is true here.  Note that in AdS the boundary terms in question live at spatial infinity, while for the Minkowski S-matrix they are defined in the asymptotic past and future. These boundary terms will be treated carefully and explicitly in the presentation that follows, where we establish, in somewhat more detail than in previous literature, the equivalence of this definition of the S-matrix with the LSZ definition, whenever both are applicable\footnote{It would be interesting to consider the extension of the AFS approach to include bound states.}.

Our perspective is that the  direct path integral definition of the S-matrix proposed by AFS is well suited to discussing matters related to asymptotic symmetries, because it is the asymptotic data that figure directly in the construction.  It is also advantageous to avoid the appearance of non-gauge invariant local correlators at an intermediate stage.

To illustrate the general approach, we first work out the example of an interacting scalar field in detail. The path integral with asymptotic boundary conditions is most efficiently evaluated by shifting the field of integration from $\phi$ to $\phi_G$ by writing  $\phi = \overline\phi + \phi_G$ where $\overline \phi$ is a   solution of the free field equation obeying the boundary conditions, so that $\phi_G$ obeys trivial asymptotic boundary conditions.  After some manipulation this leads to the following simple expression for the S-matrix operator 
\eq{i6}{  \hat{S} =  \normord{ \int\! {\cal D}\phi_G \, e^{i\int_M\!d^4x \left( {1\over 2} \phi_G \nabla^2 \phi_G  -V(\hat{\phib}+\phi_G)    \right)}    } ~.}

  The tree  approximation to the S-matrix is obtained by applying the saddle point approximation to the path integral, so that the  tree level S-matrix is given directly by the exponential of the on-shell classical action. 

We next consider scalar QED and arrive at an expression analogous to \rf{i6}, which we use to illustrate the application of this formalism to asymptotic symmetries and soft theorems, as has been the subject of much attention in recent years \cite{He:2014cra,Campiglia:2015qka,Kapec:2015ena,Kapec:2017tkm,Campiglia:2016hvg,Campiglia:2017mua,Campiglia:2018dyi}.   With the correct boundary terms in place it is easy to see that the S-matrix\footnote{We deal here with the formal, IR divergent QED  S-matrix.} is invariant under large gauge transformations that act on the asymptotic boundary conditions. This feature leads immediately to the Weinberg soft-photon theorem.  We simply need to keep track of the large gauge mode (i.e. Goldstone mode) as part of our asymptotic boundary conditions.  Although we do not work this out here, it is evident that the same approach will apply to asymptotic symmetries and soft theorems in other contexts.  

This formalism is also well suited to studying the S-matrix in curved spacetime.   As an example, we work out the case of a free scalar field defined on a general globally hyperbolic, asymptotically flat spacetime.  In the absence of  a timelike Killing vector, particles will be produced by the background geometry; the corresponding amplitudes are usually derived in the framework of  canonical quantization;  see e.g.  \cite{Birrell:1982ix,DeWitt:1975ys}.   We show that the full S-matrix, expressed in terms of the Bogoliubov coefficients relating incoming and outgoing positive frequency mode solutions,  is rather simple to derive in the AFS approach.  

In this work we often focus on  massless fields for concreteness. Only the location of the boundary terms, present at timelike infinity instead of null infinity, changes in the presence of massive fields. But since we rewrite these boundary terms in the bulk, this only changes details at intermediate steps. The ability to perform this rewriting is generic, as demonstrated in a simplified example in appendix \ref{Sec:Boundary Term Cancellation}.

The remainder of this paper is organized as follows.  In section \rff{AFS} we lay out the basic formulas for the AFS S-matrix in the context of an interacting scalar field.   In section \rff{Sec:Interacting scalar field}  we give a careful discussion of the equivalence, to all orders in perturbation theory, between the AFS and LSZ prescriptions for the S-matrix.   We turn to scalar QED in section \rff{Sec:Scalar QED}, and verify that the action is invariant under a class of large gauge transformations.    In section \rff{LGT and soft} we explain how this gauge invariance leads to a tidy  derivation of the Weinberg soft photon theorem, following  the same general philosophy as in \cite{He:2014cra}.  We generalize to curved spacetime in section \rff{Sec:Paricle production in curved space}, and show how to efficiently compute the S-matrix for a free scalar field in terms of the Bogoliubov coefficients.  We conclude with some comments in section \rff{discussion}. In appendix  \rff{GBT}   we provide more pedagogical details on the AFS approach, and also present some shortcuts for putting the action in a usable form.  Appendix \rff{Sec:Sourcing Approach} gives a slightly different approach to deriving the soft theorem, emphasizing the S-matrix generating functional rather than the S-matrix operator.  In appendix \rff{sec:AdSCFT} we derive the equivalence between the two forms of AdS boundary correlators by an argument that parallels our discussion of the S-matrix.  In appendix \rff{classical}  we work out some details regarding the form of classical solutions that govern the tree level S-matrix, emphasizing that they differ from classical solutions that are relevant for ordinary classical physics.

\section{AFS S-matrix: S-matrix as a path integral}
\label{AFS}

In this section we first review the basic setup of the AFS approach to the S-matrix. We start from the operator definition of the S-matrix (see e.g. \cite{Weinberg:1995mt}),\footnote{Here we consider a time independent Hamiltonian, though in section \rff{Sec:Paricle production in curved space} we will consider an example where this condition is relaxed.}
\eq{r1}{ \hat{S} = \lim_{\substack{t_i\rt -\infty\\t_f \rt +\infty}}e^{i\hat{H}_0 t_f} e^{-i \hat{H}(t_f-t_i) } e^{-i\hat{H}_0 t_i}~, }
where $\hat{H}_0$ denotes the free part of the full Hamiltonian $\hat{H}$.  To proceed we consider matrix elements of $\hat{S}$ in a coherent state basis, corresponding to eigenstates of annihilation operators.   For definiteness, consider a real scalar field in flat Minkowski space\footnote{Throughout, we use the mostly plus metric convention.}.  The asymptotic boundary conditions \eqref{i3} can be captured conveniently by introducing an auxiliary free field $\phib(x)$, whose mode expansion splits into positive and negative frequency components, 
\eq{r2}{ \hat{\phib}(x) = \int\frac{\dr^3p}{(2\pi)^3}\frac{1}{2\omega_p}\left(\hat b(\vec p)e^{ip\cdot x} + \hat b^\dag(\vec p)e^{-ip\cdot x}\right), \ \ \ \ p^0 = \omega_p = \sqrt{\vec{p}^2+ m^2}~, }
so
\eq{r3}{ \hat{\phib}_+(x)  = \int\! {d^3p \over (2\pi)^3} {1\over 2\omega_p}  \hat b(\vec{p}) e^{ip\cdot x},\ \ \ \ 
\hat{\phib}_-(x) =  \int\! {d^3p \over (2\pi)^3} {1\over 2\omega_p} \hat b^\dagger(\vec{p}) e^{-ip\cdot x}~. }
Throughout, we use overbars to denote free fields.   Coherent states obey    
\eq{r4}{ \hat{\phib}_+(x) | \phib_+(x)\rangle &= \phib_+(x) | \phib_+(x)\rangle \cr
\langle \phib_-(x)|\hat{\phib}_-(x) &=\langle \phib_-(x)| \phib_-(x)~. }
The inner product is given by
\eq{r4-1}{
    \langle \overline\phi_-(x)| \overline\phi_+(x)\rangle &= \exp\left[\int\frac{\dr^3p}{(2\pi)^3}\frac{1}{2\omega_p}b^\dag(\vec p)b(\vec p)\right]\cr
    &= \exp\left[ i\int_\Sigma\dr^3 x\sqrt{h}t^\mu(\overline \phi_-(x)\p_\mu \overline\phi_+(x) - \p_\mu\overline\phi_-(x)\overline\phi_+(x))\right]% = -2(\overline \phi_-, \overline\phi_+)_\Sigma~.
}
where $t^\mu$ is the future directed timelike unit normal to $\Sigma$.    Note that  $\overline \phi_-$ and $\overline\phi_+$ are independent, not necessarily related by complex conjugation.

We now define
\eq{r5}{ S[\phib] &= \langle \phib_-(\vec{x},t=0) | \hat{S} |\phib_+(\vec{x},t=0)\rangle\cr
& =  \lim_{\substack{t_i\rt -\infty\\t_f \rt +\infty}}\langle \phib_-(\vec{x},t_f) | e^{-i\hat{H}(t_f-t_i)} |\phib_+(\vec{x},t_i)\rangle ~.}
Here $\phib(x) = \phib_+(x)+ \phib_-(x)$, but we allow $\phib_\pm(x)$ to be independent, not related by complex conjugation.  The function $\phib(x)$ appearing in \rf{r5} thus represents a complex-valued free field solution which encodes the boundary conditions of the scattering problem.

As defined above, $S[\phib]$ admits a path integral representation in which the positive (negative) part of $\phi$ agrees with the corresponding part of $\phib$ in the far past (future). For a real scalar field with Hamiltonian 
\eq{r6}{ H = \int_\Sigma d^3x \left( {1\over 2} \pi_\phi^2 + {1\over 2}(\vec{\nabla} \phi)^2 +V(\phi) \right)  }
we have 
\eq{r7}{ S[\phib] = \lim_{\substack{t_i\rt -\infty\\t_f \rt +\infty}}~\int^{\phi_-(t_f) = \overline\phi_-(t_f)}_{\phi_+(t_i) = \overline\phi_+(t_i)} {\cal D}\phi \, e^{ iI[\phi,\phib]} }
where the integration is over field configurations satisfying the asymptotic boundary conditions\footnote{These boundary conditions are generated by the free evolution factors in \eqref{r1}. See Appendix \ref{Sec:QM Example} for a more detailed discussion of this point.}
\eq{r10}{ \phi_-(x) &\sim \phib_-(x)~,\quad t\rt +\infty \cr
\phi_+(x) &\sim \phib_+(x)~,\quad t\rt -\infty~. }
The action appearing in this path integral is
\eq{r8}{ I[\phi,\phib] & = \int\! d^4x \left( {1\over 2} \phi \nabla^2 \phi -V(\phi) \right) + I_\text{bndy}[\phi,\phib]  }
where  the boundary terms in the action  play an important role due to the non-vanishing boundary conditions on the fields.

The form of the required boundary terms follow from explicit construction of the phase space path integral, but can also be deduced by demanding that the total action is stationary under variations that preserve the boundary conditions\footnote{Of course, the variational principle does not fully fix the  boundary terms since one can always add  terms that depend solely on the fixed boundary data.  That no such terms should be present can be deduced in various ways: directly from the path integral construction; by demanding a trivial S-matrix in the free limit; by demanding equivalence with LSZ; or by requiring that $b^\dag(\vec p)$ to be the canonical conjugate of $b(\vec p)$, as reviewed in appendix \ref{Sec:QM Example}.   A corresponding ambiguity regarding nonlocal  boundary terms in AdS may be fixed by demanding that the boundary correlators  are compatible with the structure of a local CFT.}, closely paralleling the situation in AdS/CFT \cite{Papadimitriou:2007sj}. For this theory, the boundary terms are found to be 
\eq{r9}{
    I_\text{bndy}[\phi,\phib] &= \left( \phib_-, \phi_+ \right)_{\Sigma_f} - \left( \phib_+, \phi_- \right)_{\Sigma_i}~. }
Here  $\Sigma_i$ and $\Sigma_f$ are Cauchy surfaces, which may be thought of as constant time slices, and we have defined the scalar product
\eq{r9a}{ 
    \left( \phi, \psi \right)_\Sigma &= \frac12 \int_\Sigma\! d^3x \sqrt{h} \, t^\mu (\phi \, \p_\mu \psi - \p_\mu \phi \, \psi)~,}
where $t^\mu$ is the future-directed unit normal, e.g. $t^\mu \p_\mu = \p_t$ for equal time surfaces in Cartesian coordinates. The scalar product $\left( \phi, \psi \right)_\Sigma$, closely related to the standard Klein-Gordon inner product\footnote{We use this definition because it is what appears most naturally in these boundary terms. We will have occasion to use the standard definition of the Klein-Gordon inner product later in section \ref{Sec:Paricle production in curved space} when we consider particle production on a curved spacetime.}, vanishes between solutions which contain only the same frequency  content: $(\overline \phi_{1,+}, \overline\phi_{2,+}) = (\overline \phi_{1,-}, \overline\phi_{2,-}) = 0$. 

To extract S-matrix elements in the standard Fock space basis involving particles with definite momenta, one may take derivatives with respect to $\overline\phi_\pm$ or, more properly, the modes, to generate insertions of creation/annihilation operators before setting $\overline \phi = 0$,
\eq{ci}{
\langle q_1,\ldots, q_M|\hat S|p_1,\ldots,p_N\rangle =\left[\prod^N_{k \text{ in}}\left( {2\omega_{p_k}(2\pi)^3}\frac{\delta}{\delta b_k(\vec p_k)} \right)\prod^M_{\ell \text{ out}}\left( {2\omega_{q_\ell}}{(2\pi)^3}\frac{\delta}{\delta b^\dag_\ell(\vec q_\ell)} \right)S[\overline\phi] \right]_{\overline\phi = 0}.
}

Alternatively, standard coherent state formulas allow one to reconstruct an operator, as a normal ordered expression in terms of creation and annihilation operators,  from its matrix elements in the coherent state basis \cite{Balian:1976vq},
\eq{r11}{ \hat{S} =  \normord{ e^{-iI_\text{bndy}[\hat{\phib},\hat{\phib}]} S[\hat{\phib}] } ~.  }
In more detail, the object $\hat{\phib}$  is given by \rf{r2} and  the full expression on the right hand side of \rf{r11} is normal ordered with respect to the operators $\hat b(\vec{p}), \hat b^\dagger(\vec{p}) $. Note that the $e^{-iI_\text{bndy}[\hat\phib, \hat \phib]}$ factor does not completely cancel the boundary term, but only the free part. In this paper, we primarily work with the normal ordered form of the S-matrix rather than its generating functional form, but in appendix \ref{Sec:QM Example}, we review both the generating functional approach and the derivation of the identity leading to \eqref{r11}.

The above expressions are valid to all orders of perturbation theory (though if massless particles are present the expressions may of course be infrared divergent).   If we restrict to tree level then \rf{r7} gives a direct relation between the S-matrix generating functional and the classical action, $S_\text{tree}[\phib] = e^{i I[\phi_{cl},\phib]}$, where $\phi_{cl}$ is the classical solution of the field equations  that obeys the boundary conditions $\phib$, as we will see in more detail later.

As a final point in this section, since the boundary conditions allow $b(\vec p)$ and $b^\dag(\vec p)$ to be independent, not related by complex conjugation, the prescription for computing the S-matrix will involve complex configurations of the real scalar field $\phi(x)$. This can be understood as corresponding to the fact that off-diagonal matrix elements $\langle\psi_\text{out}|\hat\phi(x)|\psi_\text{in}\rangle$ are in general complex, even for a Hermitian field operator $\hat\phi(x)$.   The tree level classical solutions also have different asymptotics than one might expect from purely classical considerations; this is discussed in more detail in  appendix \rff{classical}.

\section{Equivalence with the LSZ prescription} \label{sec: int scalars}\label{Sec:Interacting scalar field}

It is instructive to verify in detail the perturbative equivalence of the AFS and LSZ prescriptions for the S-matrix, which in particular involves keeping careful track of the boundary terms, which were left largely implicit in much of the previous literature on this topic.  The strategy will be to consider the path integral in the presence of both bulk sources and boundary conditions, allowing us to toggle back and forth between the AFS and LSZ prescriptions. 

Part of our procedure will involve using Stokes' theorem to  rewrite boundary terms as bulk terms, which leads to some cancellation.   We work through this very explicitly, but it is useful to note that for the purpose of more efficient computation there is a quicker way to arrive at the final result for the scattering action, as discussed in appendix  \rff{GBT}.

\subsection{Action and boundary terms}

As in the previous section, we consider the example of a real scalar field, but now take the field to be massless for concreteness,\footnote{As far as this section is concerned, including masses just means considering asymptotic data on $i^\pm$ rather than ${\cal I}^\pm$. }  and we now turn on a bulk  source $J(x)$. The action is then\footnote{Here and in what follows we suppress the volume element $\sqrt{-g}$ in all $d^4x$ integrals.} 
\eq{bk}{ I &=   \int_M\! d^4x  \big( {1\over 2} \phi\nabla^2 \phi-V(\phi)+J\phi \big)  +I_\text{ct} +I_\text{bndy} \cr& = I_\text{bulk} +I_\text{ct} +I_\text{bndy}~ }
where we assume $V(0)=V'(0)=V''(0)=0$. Here $I_{ct}$ denotes the counterterms required to cancel UV divergences; it will mostly be suppressed in what follows.

The integration region $M$ is bounded in the past and future by initial and final surfaces $(\Sigma_i,\Sigma_f)$. These surfaces are eventually taken to approach $(\scrI^-,\scrI^+)$.    To implement this  we introduce retarded and advanced coordinates, $(r,u,x^A)$ and $(r,v,x^A)$ where $u = t - r$, $v = t+ r$, and $x^A$ are the coordinates on the unit 2-sphere. The Minkowski metric becomes
\eq{l10}{ ds^2 
&= -du^2 -2dudr +r^2\dr\Omega^2\cr
&= -dv^2 +2dvdr +r^2\dr\Omega^2~.}
For $\Sigma_f$ we can then consider a 1-parameter family of spacelike hypersurfaces defined by $F_f^\alpha(u,r,x^A)=0$ such that, as $\alpha \to 1$ from below, solutions at fixed $(u,x^A)$ send $r\to \infty$, and such that the normal vector $t_\mu = \nabla_\mu F_f^\alpha$ approaches $t^\mu \p_\mu = \p_u$ as $\alpha \to 1$.  The initial surface $\Sigma_i$ is treated analogously, now with $t^\mu \p_\mu \to \p_v$ as $\alpha \to -1$.   At intermediate values of $\alpha$ the normal vector $t^\mu$  is taken to be timelike and pointing towards the future.

We fix $I_\text{bndy}$ by demanding a good variational principle compatible with the boundary conditions \rf{r10} of the path integral.  To that end we note that the variation of the action is 
 \eq{bl}{ \delta I &= \int_M \! d^4x  E[\phi]\delta \phi - (\phib_-,\delta  \phi_+ )_{\Sigma_f} + (\phib_+,\delta  \phi_- )_{\Sigma_i}  + \delta I_\text{bndy} ~.}
Here we are considering variations that preserve the boundary conditions, and the Euler-Lagrange equations are 
\eq{bm}{ E[\phi] = \nabla^2 \phi -V'(\phi) + J=0~. }
Demanding that $\delta I=0$ when $E[\phi]=0$ fixes the boundary term to be
\eq{bn}{  I_\text{bndy} = (\phib_-,\phi_+)_{\Sigma_f} - (\phib_+,\phi_-)_{\Sigma_i}~.}

\subsection{Rewriting the action}

Our next task is to rewrite the action in a form such that all dependence on the boundary conditions $\phib$ and source $J$ is via the function $\phib_J(x)$, which is defined to obey
\eq{bo}{\nabla^2 \phib_J(x) +J(x)=0   }
as well as the boundary conditions \rf{r10}, 
\eq{bp}{ \phib_{J,-}(x) & = \phib_-(x) \quad {\rm on~}\scrI^+ \cr
\phib_{J,+}(x) & = \phib_+(x) \quad {\rm on~}\scrI^-}
This solution to the free, inhomogeneous equation of motion can be constructed from a free solution to the homogeneous problem and the Feynman Green's function (propagator) $G_F(x)$, which obeys $\nabla^2G_F(x) = i\delta^{(4)}(x)$ and is purely positive (negative) frequency for $x^0$ positive (negative). The inhomogeneous solution is then given by
\eq{bq}{ \phib_J(x) = \phib(x)+i\int\! d^4y G_F(x-y)J(y)~.}
A general off-shell field configuration obeying the boundary conditions will then be written as
\eq{br}{ \phi(x) = \phib_J(x) + \phi_G(x) }
where $\phi_G$ obeys a falloff that respects the boundary condition \rf{r10} on $\phi$.

To rewrite the action in terms of $\phib_J$, it will prove useful to reexpress $I_\text{bndy}$ as a bulk integral.
We first of all use the stated boundary conditions to write
\eq{bt}{ I_\text{bndy} = I_\text{bndy}^\text{free} + I_\text{bndy}^G }
where
\eq{bu}{ I_\text{bndy}^\text{free} & = (\phib_{J,-},\phib_{J,+})_{\Sigma_f} - (\phib_{J,+},\phib_{J,-})_{\Sigma_i}~,\cr
I_\text{bndy}^G  & = (\phib_{J,-},\phi_{G})_{\Sigma_f} - (\phib_{J,+},\phi_{G})_{\Sigma_i}~.}
Next, we note
\eq{bw}{ \nabla^\mu ( \phib_J \p_\mu \phi_G - \p_\mu \phib_J \phi_G) & = \phib_J \nabla^2 \phi_G - \nabla^2 \phib_J \phi_G}
which allows us to write 
\eq{bx}{ I_\text{bndy}^G  & = -{1\over 2} \int_M\! d^4x  \big(\phib_J \nabla^2 \phi_G -\nabla^2 \phib_J\phi_G \big)~. }
Turning now to $I_\text{bulk} $, writing  $\nabla^2 \phi = \nabla^2\phib_J+\nabla^2 \phi_G$ and $J=-\nabla^2 \phib_J$ we have 
\eq{by}{ I_\text{bulk} = \int_M\! d^4x  \left( {1\over 2} \phi \nabla^2 \phi_G  -V(\phi) -{1\over 2}\nabla^2 \phib_J \, \phi  \right)~.  }
This gives
\eq{bz}{ I_\text{bulk} + I_\text{bndy}^G = \int_M\! d^4x \left( {1\over 2} \phi_G \nabla^2 \phi_G  -V(\phib_J+\phi_G) -{1\over 2}\nabla^2 \phib_J \, \phib_J  \right)  ~. }

\subsection{Equivalent forms of the S-matrix  }

The total action  separates into two pieces
\eq{ca}{ I = I_0[\phib_J] + I_G[\phib_J,\phi_G] }
with 
\eq{cb}{ I_0[\phib_J] &=I_\text{bndy}^\text{free}[\phib_J] -{1\over 2}  \int_M\! d^4x  \nabla^2 \phib_J \phib_J,\cr
I_G[\phib_J,\phi_G]&=  \int_M\! d^4x  \left( {1\over 2} \phi_G \nabla^2 \phi_G  -V(\phib_J+\phi_G)  \right)~.
}
Written in this form, the total action  has two key properties.  First, there is no explicit dependence on $\phib$ or $J$; all such dependence has been absorbed into the function $\phib_J(x)$.  Second, in  $I_G[\phib_J,\phi_G]$ which contains all the dependence on interactions, $\phib_J$ enters without any derivatives. 

We now consider the path integral in the presence of source $J$ and  boundary conditions specified by $\phib$,
\eq{cd}{ Z[\phib,J] = e^{i I_0[\phib_J] } \int\! {\cal D} \phi_G \, e^{i I_G[\phib_J,\phi_G]}~.}

We first recall the LSZ prescription for the S-matrix.   In this case we set $\phib=0$ and expand $Z[0,J]$ as
\eq{ce}{ Z[0,J] = 1+\sum_{n=1}^\infty{i^n \over n!}\int\! \left[\prod_{i=1}^n d^4x_i d^4 y_i\right]  G_\text{amp} ^{(n)}(x_i)G_F(x_1-y_1)J(y_1)\ldots G_F(x_n-y_n)J(y_n)   }
where $G_\text{amp}^{(n)}(x_i)$ are the (generally disconnected) $n$-point amputated correlators.
The S-matrix generating functional $S[\phib]$ is obtained by stripping off the external sources and propagators and replacing them by on-shell wavefunctions, an operation which amounts to\footnote{Here we assumed a renormalization prescription that sets the residue of the pole in the two-point function to unity; see below.}
\eq{cg}{ i G_F J \equiv \int\! d^4y \, iG_F(x-y)J(y) \, \rt \, \phib(x) }
where $\phib(x)$ is given by the mode expansion \rf{r2}.  We thus write
\eq{ch}{ S_{\rm LSZ}[\phib] = Z[0,J]|_{iG_FJ ~\rt ~\phib}~. }
A specific S-matrix element is then obtained by differentiation, as in \rf{ci}.

Turning now to the AFS prescription, we consider the path integral in the absence of sources but with boundary condition specified by $\phib$; this quantity is by definition $Z[\phib,0]$, so 
\eq{chy}{ S_{\rm AFS}[\phib] = Z[\phib,0]~.}
To prove that this is equal to the LSZ S-matrix generating functional, i.e that $ S_{\rm AFS}[\phib] =  S_{\rm LSZ}[\phib]$, we need to establish the relation
\eq{ck}{ Z[\phib,0] = Z[0,J]|_{iG_FJ ~\rt ~\phib}~. }
A naive argument to this effect is that $Z[\phib,J]$ in \rf{cd} only depends on $(\phib,J)$ via $\overline \phi_J$, and since $ \overline\phi_J(x) = \phib(x)+i\int\! d^4y G_F(x-y)J(y)$ it is apparent that the substitution in question leaves $\phib_J$ invariant.  This is too quick: the combination $\phib_J +iG_F\nabla^2 \phib_J$ vanishes for any $J$ when $\phib=0$, so the $\phib_J$ dependence of $Z$ cannot be uniquely reconstructed, without further input, just from the $J$ dependence alone.  The missing ingredient is obtained by considering the diagrammatic computation of \rf{cd} with $\phib=0$ and $J\neq 0$.  The position space Feynman diagrams will involve some combination of  $J$'s, $\phib_J$'s and $G_F$'s integrated against each other; importantly, since $\phib_J$ appears in  $I_G$ without derivatives, the Feynman integrands involve no derivatives of $\phib_J$.  On the other hand, the ambiguity involving the addition of  $\phib_J +iG_F\nabla^2 \phib_J$ does involve such derivatives. So once we stipulate that $Z$ can be written in such a non-derivative form then we see that the naive substitution rule $iG_F J \rt \phib$ is indeed valid. From this we conclude that the LSZ S-matrix generating function is equal to the path integral in the absence of sources, and thus that 
\eq{cl}{  S_{\rm AFS}[\phib] =  S_{\rm LSZ}[\phib]~.}

Focusing now on the AFS version, we note that with the sources turned off we can simplify the action. Since $\phib_J=\phib$ and  $\nabla^2\phib   =0$ when $J=0$ we have
\eq{cla}{ I_0[\phib] & = I_\text{bndy}^\text{free}[\phib_J]~.}
We also note that from \rf{bu} when $J=0$ the two terms in $I_\text{bndy}^\text{free} $ are equal  because the scalar inner product is preserved under free time evolution. We then  have 
\eq{cm}{ S[\phib]= Z[\phib,0] =e^{iI_0[\phib]}\int\! {\cal D}\phi_G e^{i\int_M\!d^4x \left( {1\over 2} \phi_G \nabla^2 \phi_G  -V(\phib+\phi_G)    \right)  +iI_{ct}}~, }
where we restored the counterterm action.   Regarding the counterterms, the consistency of the above discussion requires a renormalization prescription such that the sum of self-energy diagrams obeys  $\Sigma(0)= \Sigma'(0)=0$, where the former implies that the physical particle mass is $m^2=0$ and the latter implies that the residue at the pole is unity.

To get the S-matrix operator we strip off the prefactor in \rf{cm}, as explained in appendix \ref{GBT}, and interpret what remains as a normal ordered operator expression, 
\eq{cn}{  \hat{S} =  \normord{ \int\! {\cal D}\phi_G \, e^{i\int_M\!d^4x \left( {1\over 2} \phi_G \nabla^2 \phi_G  -V(\hat{\phib}+\phi_G)    \right)  +iI_{ct}}  }~. }

The result \rf{cn} holds to all orders in perturbation theory.  At tree level one should solve $\nabla^2 \phi_G = V'(\phib+\phi_G)$ and then evaluate $\hat{S}$ at this saddle point, yielding 
\eq{co}{ \hat{S}_{\rm tree} =  \normord{  e^{i\int_M\!d^4x \left( {1\over 2} \phi_G V'(\hat{\phib}+\phi_G)    -V(\hat{\phib}+\phi_G)    \right) }  }~,}
where $\phi_G$ is computed order by order using the Feynman propagator, e.g., 
\eq{cp}{\phi_G(x) =i\int\! d^4y \, G_F(x,y) V'\big(\phib(y)\big) + {\cal O}(V^2)~.  }
It is straightforward to verify that this reproduces the standard expansion in terms of Feynman diagrams. 

\section{Scalar QED}\label{Sec:Scalar QED}

We now consider a massless complex scalar field coupled to electromagnetism, with action 
\eq{s1}{ I  &=  \int\! d^4x  \left( {1\over 2} A^\mu \nabla^2 A_\mu +{1\over 2}\phi^* D^2 \phi + {1\over 2} (D^2 \phi)^*  \phi + J^*\phi + J \phi^* - J^\mu A_\mu \right)\cr
&\quad + I_\text{ct}+I_\text{ghosts}+I_\text{bndy}~.  }
The counterterm and ghost actions will be suppressed.     The covariant derivative is
\eq{s2}{ D_\mu \phi = (\p_\mu -i eA_\mu)\phi}
corresponding to the gauge invariance
\eq{s3}{ \phi \rt e^{ie\lambda }\phi~,\quad A_\mu \rt A_\mu +\p_\mu \lambda~. }
We work in Lorenz gauge,
\eq{s4}{ \nabla^\mu A_\mu=0,}
and the gauge fixing term in \rf{s1} was chosen to make the gauge kinetic term take the form $ {1\over 2} A^\mu \nabla^2 A_\mu$. There are residual gauge transformations preserving Lorenz gauge which satisfy $\nabla^2 \lambda = 0$; these will play an important role in the derivation of the soft theorem.

We now manipulate the action in a way that directly parallels what was done in the scalar field example, so we will suppress some details.   Boundary conditions on the positive (negative) frequency components of $(\phi,A_\mu)$ are imposed just as before, and the associated boundary action is
\eq{s5}{ I_\text{bndy} &= (\Ab^\nu_-,A_{\nu+})_{\Sigma_f} + (\phib^*_-,\phi_+)_{\Sigma_f} + (\phib_-,\phi^*_+)_{\Sigma_f} \cr
&  \quad -(\Ab^\nu_+,A_{\nu-})_{\Sigma_i}  - (\phib^*_+,\phi_-)_{\Sigma_i} - (\phib_+,\phi^*_-)_{\Sigma_i}~.}
 One point worth noting is that the ``Cartesian components'' of $A_\mu$, meaning the components in coordinates where the metric takes the form $\dr s^2 = -\dr t^2 + \dr \vec x^2$, are taken to vanish as $r\rt \infty$.  We used this assumption, together with the corresponding fall-off of the scalar field, to replace covariant derivatives in the boundary term by ordinary derivatives, since the difference vanishes as $r\rt \infty$.

We now write
\eq{s6}{ \phi & = \phib_J + \phi_G \cr
\phi^* & = \phib^*_J + \phi^*_G \cr
A^\mu & = \Ab^\mu_J + A^\mu_G }
with
\eq{s7}{ \phib_J(x) & =  \phib(x)+ i\int\! dy \, G_F(x,y)J(y) \cr
 \phib^*_J(x) & =  \phib^*(x)+ i\int\! dy \, G_F(x,y)J^*(y) \cr
 \Ab^\mu_J(x) & = \Ab^\mu(x) - i\int\! dy \, G_F(x,y) J^\mu(y) }
where $\nabla^2 \phib=\nabla^2 \phib^* = \nabla^2 \Ab^\mu=0$.  Inserting \rf{s6} into the boundary term, rewriting the $G$-dependent part as a bulk integral, and trading away sources using $J=-  \nabla^2 \phib_J $ etc, 
we arrive at 
\eq{s8}{ I = I_0[\phib_J,\phib_J^*,\Ab^\mu_J]  + I_G[\phib_J,\phib_J^*,\Ab^\mu_J; \phi_G,\phi_G^*,A^\mu_G  ]  }
with
\eq{s9}{ I_0 &= (\Ab^\nu_{J-} ,\nabla_\mu \Ab_{J\nu+})_{\Sigma_f} + (\phib^*_{J-},\phib_{J+})_{\Sigma_f} + (\phib_{J-},\phib^*_{J+})_{\Sigma_f} \cr
& \quad  -(\Ab^\nu_{J+} ,\nabla_\mu \Ab_{J\nu-})_{\Sigma_i} -(\phib^*_{J+},\phib_{J-})_{\Sigma_i} -  (\phib_{J+},\phib^*_{J-} )_{\Sigma_i}  \cr
& \quad   - {1\over 2} \int\! d^4x \left(  \phib_J\nabla^2 \phib_J^*  +   \phib_J^*\nabla^2 \phib_J +  \Ab_{J\mu}\nabla^2 \Ab_J^\mu   \right)  }
and 
\eq{s10}{  I_G & =    {1\over 2} \int\! d^4x \left(  A_G^\mu \nabla^2 A^G_\mu  -\phib_J^* \nabla^2 \phi_G -\phib_J \nabla^2 \phi^*_G  \right) \cr
& \quad +{1\over 2} \int\! d^4x \left(  \phi^* [ D^2 \phi- \nabla^2 \phib_J]+ \phi   [ (D^2 \phi)^*-\nabla^2 \phib_J^*] \right)~.  }
At this point we can make essentially the same argument as in the scalar case to argue that the LSZ prescription applied to the path integral with vanishing boundary conditions is equivalent to computing the source free path integral with nontrivial boundary conditions.   The only slight subtlety comes from the fact that $\nabla^2 \phib_J$ appears in $I_G$, so we need to revisit the ambiguity under the addition of  $\phib_J +iG_F\nabla^2 \phib_J$ when we try to reconstruct the full $\phi_J$ dependence from the $\phib=0$ result.  However, we have written \rf{s10} to make explicit the fact that no second derivatives acting on $\phib_J$ appear, due to the cancellation between $D^2$ and $\nabla^2$.  This lack of second derivatives removes the ambiguity, just as in the pure scalar case. 

Having established the equivalence with LSZ,  to obtain the final expression for the AFS S-matrix generating functional we turn off the sources, which implies $\phib_J  = \phib$ so $\nabla^2 \phib_J = \nabla^2 \phib_J^* = \nabla^2 \Ab_J^\mu=0$, yielding 
 \eq{s11}{  I_G=   {1\over 2} \int\! d^4x \left(  A_G^\mu \nabla^2 A^G_\mu  -\phib^* \nabla^2 \phi_G -\phib \nabla^2 \phi^*_G + \phi^*  D^2 \phi+  (D^2 \phi)^* \phi \right)~   }
with $\phi= \phib+\phi_G$, $\phi^*= \phib^*+\phi^*_G$, $A^\mu= \Ab^\mu+A^\mu_G$.
The S-matrix generating functional is then 
\eq{s12}{ S[\phib,\phib^*,\Ab_\mu] & = e^{i I_0[\phib,\phib^*,\Ab_\mu]} \int\! {\cal D} \phi_G{\cal D} \phi^*_G {\cal D}A^\mu_G~e^{iI_G } }
and the S-matrix operator is 
\eq{s13}{ \hat{S} = \normord{ \int\! {\cal D} \phi_G{\cal D} \phi^*_G {\cal D} A^\mu_G~e^{iI_G }   }~. }

\subsection{Generalization}

In arriving at \rf{s11} we assumed that the barred fields, which encode the boundary data, obeyed free field equations, $\nabla^2 \phib = \nabla^2 \phib^* = \nabla^2 \Ab_\mu =0$.  This is convenient for computational purposes, but we should recall that since the S-matrix generating functional only depends on asymptotic boundary data, we are free to deform the barred fields provided that we keep fixed their leading asymptotic form.  This freedom will be important for the discussion of gauge invariance in the next section, since the  gauge transformation $\phib(x) \rt e^{ie\lambda(x)} \phib(x)$ does not preserve the equation $\nabla^2 \phib=0$.    

To this end we write down the generalization of \rf{s11} for the larger space of barred fields.   In writing \rf{s11} we chose to express all boundary terms as bulk terms, but for the purposes of establishing gauge invariance it is more convenient to leave some scalar contributions in the form of boundary terms.   This yields
\eq{s11zz}{  I_G& =  {1\over 2}\int\! d^4x \Big(  A_G^\mu \nabla^2 A^G_\mu +(D^2 \phi)^* \phi + \phi^*D^2\phi\Big)  \cr
& \quad +   (\phib^* , \phi_G)_{\Sigma_f}+ (\phib, \phi_G^*)_{\Sigma_f} -   (\phib^* , \phi_G)_{\Sigma_i}- (\phib, \phi_G^*)_{\Sigma_i} }
As explained above, for fixed asymptotic data this version of the action will produce the same S-matrix generating function as \rf{s11}, but it holds for a larger space of barred fields $(\phib,\phib^*,\Ab_\mu)$, namely those that do not obey free field equations. 

\subsection{Gauge invariance}

We now demonstrate  invariance of the S-matrix under a class of large gauge transformations.    As a reminder,  the sources are set  to zero, $J=J^*= J^\mu=0$, so that $\phib_J=\phib$ and $\Ab_J=\Ab $.

We take a gauge transformation to act as 
\begin{alignat}{2}
\label{s14}
    \phib   &\to e^{ie\lambda}\phib~,\quad        &  \phi_G   &\to e^{ie\lambda}\phi_G,  \nonumber \\
    \phib^* & \to e^{-ie\lambda}\phib^*~,\quad    &  \phi^*_G &\to e^{-ie\lambda}\phi^*_G,  \nonumber \\
    \Ab_\mu &\to \Ab_\mu + \p_\mu \lambda~,\quad  &  A^G_\mu  &\to A^G_\mu~.
\end{alignat}
The transformation of $(\phi_G,\phi_G^*)$ should be thought of as a change of integration variable.
Holding either $u$ or $v$ fixed, we allow for the residual transformations $\lambda(x)$, obeying $\nabla^2\lambda(x) = 0$, to have nonzero large $r$ limits, meaning $\lambda(x)$ may be nonzero on $\scrI^\pm$; however, we demand that the limiting value be independent of $(u,v)$ in order that $(A_u,A_v)$ vanish as $r\rt \infty$.  Under these conditions $\lambda(x)$ obeys an antipodal relation \cite{Campiglia:2017mua} that equates its value at a point on $\scrI^+$ to its value at the antipodal point on $\scrI^-$.  To summarize,
\eq{s15}{  \lim_{r\rt \infty} \lambda(x)  = \left\{ \begin{array}{cc}
  \lambda_0(x^A) & {\rm on~}\scrI^+ \\
\lambda_0(x'^A) & {\rm on~}\scrI^- 
\end{array} \right. }
where $x^A \rt x'^A$ is the antipodal map on $S^2$.  In \rf{s11zz} the terms in the top line are manifestly gauge invariant.  As for the boundary terms in the second line, using that $n^\mu \p_\mu $ is equal to $-\p_u$ on $\scrI^+$ and $\p_v$ on $\scrI^-$,  we see that the $(u,v)$ independence of $\lambda_0$ implies invariance of these terms. 
%In \rf{s11} the only terms that are not manifestly gauge invariant are $\phib^* \nabla^2 \phi_G$ and $\phib \nabla^2 \phi^*_G$.   However, upon writing 
%
%\eq{s16}{  \int_M\! d^4x \sqrt{g} \phib^* \nabla^2 \phi_G &=
%\int_{\scrI^+ \cup \scrI^-}  \!  du d^2x \sqrt{\gamma} r^2 n^\mu [ %\phib^* \nabla_\mu \phi_G - \nabla_\mu \phib^* \phi_G  ]     }
%
%and using that $n^\mu \p_\mu $ is equal to $-\p_u$ on $\scrI^+$ and $\p_v$ on $\scrI^-$,  we see that the $(u,v)$ independence of $\lambda_0$ implies invariance of this term, and likewise for  $\phib \nabla^2 \phi^*_G$.  Similarly, $I_0$ is also invariant due to the $(u,v) $ independence of $\lambda_0$.  
We conclude that
\eq{s17}{ S[ e^{ie\lambda}\phib, e^{-ie\lambda}\phib^*,\Ab_\mu+ \p_\mu \lambda]= S[\phib,\phib^*,\Ab_\mu]~,  }
and similarly for $\hat{S}$. 

Suppose that we used \rf{s11} to compute the S-matrix generating functional, taking the barred fields to obey free field equations.  In this case \rf{s17} will not hold because the gauge transformation does not respect the free field equations obeyed by the barred fields.  However, if we recast the transformation as acting on the asymptotic data, which is all that the generating functional depends on, then invariance is maintained.   In particular, consider the scalar field $\phib$ whose asymptotic form near $\scrI^+$  we write as $r^{-1} \phib_{-1}(u,x^A)$.     Under the gauge transformation we have $\phib_{-1}(u,x^A) \rt e^{ie\lambda_0(x^A)} \phib_{-1}(u,x^A) $.  Rather than thinking of the gauge transformation acting on the full bulk field $\phib(x)$ as in  \rf{s14}, we can view it as acting to transform the free field $\phib(x)$ with asymptotic data $\phib_{-1}$ to the new free field $\phib_{\lambda}$ with  asymptotic data $ e^{ie\lambda_0(x^A)}\phib_{-1}$.  This will act on the mode coefficients of $\phib$ as in 
\rf{l12} below.  An analogous statement holds for $\phib^*$ and $\Ab_\mu$.    The point to be made here is that the generating function computed using \rf{s11} is invariant under these transformations, because it defines the same transformation on the asymptotic data as the transformation \rf{s14}, for which we have already established invariance.

\section{Large gauge transformations and soft theorems}\label{LGT and soft}

We now rederive, in our language, the result of \cite{He:2014laa,He:2014cra,Strominger:2017zoo} (see also \cite{Miller:2021hty} for a nice pedagogical review)  that soft theorems follow from asymptotic symmetries.   Following \cite{He:2014cra} we start by explicitly separating out the pure gauge part of the boundary conditions on the vector potential by writing 
\eq{l1}{ \Ab_\mu = \tilde{\Ab}_\mu + {1\over e} \p_\mu \Phi}
with $\nabla^\mu\tilde{\Ab}_\mu =\nabla^2 \Phi =0 $.   The gauge mode $\Phi $ has leading $r^0$ behavior near $\scrI^\pm$,
\eq{l2}{ \Phi(x) \rt \left\{\begin{array}{cc}
  \Phi_0(x^A) & {\rm on~}\scrI^+  \\
  \Phi_0(x'^A) & {\rm on~}\scrI^- 
\end{array}  \right.  }
where $x^A \rt x'^A$ is the antipodal map on $S^2$. Thus $\Phi_0$ encodes the constant modes (with respect to $(u,v)$) of $\Ab_A$.  Hence $\tilde{\Ab}_A$ is taken to have no constant mode on $\scrI^\pm$, and as such admits a well defined Fourier transform. 

We now write the S-matrix functional as 
\eq{l3}{ S[\phib,\phib^*,\tilde{\Ab}_\mu,\Phi]~,}
and the gauge invariance statement \rf{s17} now reads  
\eq{l4}{ S[ e^{ie\lambda}\phib, e^{-ie\lambda}\phib^*,\tilde{\Ab}_\mu,\Phi+ e \lambda]= S[\phib,\phib^*,\tilde{\Ab}_\mu,\Phi]~.  }
Setting $e\lambda =-\Phi$ we rewrite this as 
\eq{l5}{ S[ e^{-i\Phi}\phib, e^{i\Phi}\phib^*,\tilde{\Ab}_\mu,0]= S[\phib,\phib^*,\tilde{\Ab}_\mu,\Phi]~.  }

In order to work with asymptotic quantities we use the asymptotic expressions for the fields, as obtained from a saddle point approximation.   The mode expansion of the complex scalar field is 
\eq{l6}{ \phib(x)  & = \int\! {d^3p \over (2\pi)^3} {1\over 2\omega_p} \left( b(\vec{p}) e^{ip\cdot x}+c^\dagger(\vec{p}) e^{-ip\cdot x}\right),\cr
\phib^*(x)  & = \int\! {d^3p \over (2\pi)^3} {1\over 2\omega_p} \left( c(\vec{p}) e^{ip\cdot x}+b^\dagger(\vec{p}) e^{-ip\cdot x}\right)~.}
The large $r$ asymptotics are
\eq{l7}{ \phib(x) \approx \left\{ \begin{array}{cc}
   {-i\over 8\pi^2 r} \int_0^\infty \! d\omega  \left( b(\omega \xh) e^{-i\omega u} -  c^\dagger(\omega \xh) e^{i\omega u}  \right) & {\rm on}~ \scrI^+   \\  & \\
   {i\over 8\pi^2 r} \int_0^\infty \! d\omega \left( b(-\omega \xh) e^{-i\omega v} -  c^\dagger(-\omega \xh) e^{i\omega v} \right) & {\rm on}~ \scrI^-  
\end{array} \right.  }
and
\eq{l8}{ \phib^*(x) \approx \left\{ \begin{array}{cc}
   {-i\over 8\pi^2 r} \int_0^\infty \! d\omega  \left( c(\omega \xh) e^{-i\omega u} -  b^\dagger(\omega \xh) e^{i\omega u}  \right) & {\rm on}~ \scrI^+   \\  & \\
   {i\over 8\pi^2 r} \int_0^\infty \! d\omega \left( c(-\omega \xh) e^{-i\omega v} -  b^\dagger(-\omega \xh) e^{i\omega v} \right) & {\rm on}~ \scrI^-  
\end{array} \right.  }
As in \cite{He:2014cra} it is useful to employ complex coordinates on the sphere,
\eq{l9}{x^1+ix^2 = {2rz\over 1+z\zb}~,\quad x^3 = r {1-z\zb \over 1+z\zb }, }
so that the sphere metric in \eqref{l10} becomes $\dr\Omega^2 = 2\gamma_{z\overline z}\dr z \dr\overline z$ with 
\eq{l11}{ \gamma_{z\zb} = {2\over (1+z\zb)^2}~.}
Using this, the transformation appearing in \rf{l5}, $\phib\rt e^{-i\Phi}\phib $ and $\phib^* \rt e^{i\Phi}\phib^*$, implies the following transformation of the modes,
\eq{l12}{ & b(\pv) \rt e^{-i\Phi_0(z,\zb) }b(\pv)~,\quad  b^\dagger(\pv) \rt e^{i\Phi_0(z,\zb) }b^\dagger (\pv) \cr
 & c(\pv) \rt e^{i\Phi_0(z,\zb) }c(\pv)~,\quad  c^\dagger(\pv) \rt e^{-i\Phi_0(z,\zb) }c^\dagger (\pv)~, }
 where the vector $\pv$ defines a point on the sphere via $\pv = \omega_p \hat{x}.$  So in terms of the modes, the statement of invariance of the S-matrix under large gauge transformations implies 
\eq{l13}{& S[  e^{-i\Phi_0(z,\zb) }b(\pv)  ,e^{i\Phi_0(z,\zb) }b^\dagger(\pv), e^{i\Phi_0(z,\zb) }c(\pv),e^{-i\Phi_0(z,\zb) }c^\dagger (\pv)  ,\tilde{\Ab}_\mu,0]\cr
& \quad\quad= S[b(\pv)  ,b^\dagger(\pv),c(\pv),c^\dagger (\pv),\tilde{\Ab}_\mu,\Phi_0] }
where we have suppressed writing the mode expansion of the gauge field.  This relation holds equally well for the S-matrix generating functional or the S-matrix operator.

The dependence \eqref{l13} on the large gauge transformations is exactly what is needed to reproduce the soft photon theorem. To show this, we introduce the operator $\hat{N}(z,\zb)$, which is conjugate to $\hat{\Phi}_0(z,\zb)$ in the sense that\footnote{These commutators follow from the symplectic form of Maxwell theory.  More complicated theories such as gravity may contain a richer set of large gauge modes and a correspondingly richer boundary symplectic form; the latter can be computed using the methods in \cite{Kim:2023vbj}, for example.   }
\eq{l14}{   [ \p_w \hat{N}(w,\wb),\hat{\Phi}_0(z,\zb) ] & = {i\over 4\pi} {1\over w-z}\cr
 [ \p_{\wb} \hat{N}(w,\wb),\hat{\Phi}_0(z,\zb) ] & = {i\over 4\pi} {1\over \wb-\zb}~. }
As discussed in \cite{He:2014cra},  the $\hat{N}$ operator is obtained by integrating the angular part of the field strength over null infinity (at fixed angular location),
\eq{l15}{ e \p_z \hat{N}(z,\zb) = \int_{I^+}\! du \hat{\overline{F}}_{uz}~ }
and creates soft photons in the sense, 
\eq{l20}{ e \p_z \hat{N} = -{1\over 8\pi^2 }{\sqrt{2} \over 1+z\zb} \lim_{\omega \rt 0^+}    [ \omega {\hat{a}}^\text{out}_+(\omega \xh) +  \omega  \hat{a}^{\dagger {\rm out}}_-   (\omega \xh)   ]~, }
where we follow the notation of \cite{Strominger:2017zoo}.

Of direct relevance here are the commutation relations
\eq{l16}{  [\p_w \hat{N}(w,\wb),e^{-iQ\hat{\Phi}_0(z,\zb)} ]&= {1\over 4\pi}{Q\over w-z} e^{-iQ\hat{\Phi}_0(z,\zb)}, \cr
 [\p_{\wb} \hat{N}(w,\wb),e^{-iQ\hat{\Phi}_0(z,\zb)} ]&= {1\over 4\pi}{Q\over \wb-\zb} e^{-iQ\hat{\Phi}_0(z,\zb)}, }
while $\hat{N}$ commutes with the other fields, $(\phib,\phib^*,\tilde{\Ab}_\mu)$. The soft theorem corresponds to evaluating the commutator of $\p_w \hat{N}$ with the S-matrix operator, which we identify with the right hand side  of \rf{l13},
\eq{l17}{ \hat{S} = \hat{S}[b(\pv)  ,b^\dagger(\pv),c(\pv),c^\dagger (\pv),\tilde{\Ab}_\mu,\Phi_0] }
where all objects appearing are now operators.
To compute the commutator $[\p_w \hat{N},\hat{S}] $ we replace $\hat{S}$ by the left hand side of \rf{l13} and use \rf{l16}.   Generalizing to the case where we have a collection of scalar fields with charges $(Q_1, Q_2, \ldots)$, we evaluate the commutator between states of definite particle number. The contributing term in the S-matrix operator then has a corresponding matched set of creation and annihilation operators upon which the commutator acts as in \rf{l16}.   This gives 
 \eq{l18}{  \langle {\rm out} |  [\p_z \hat{N}(z,\zb),\hat{S}] |{\rm in}\rangle =  {1\over 4\pi} \left[ \sum_{k~{\rm in}} {Q^{\rm in}_k \over z-z^{\rm in}_k}  -\sum_{k~ {\rm out}} {Q^ {\rm out}_k \over z-z^ {\rm out}_k}  \right] \langle  {\rm out}|\hat{S}|{\rm in}\rangle~, }
which agrees with the statement of the soft theorem in section 2.8.3 of \cite{Strominger:2017zoo}.

In more detail, the celestial sphere locations $(z,z_k^{\rm in},z_k^{\rm out})$ may be converted into the momenta of the particles, along with the two transverse polarizations of the soft photon; see \cite{Strominger:2017zoo}.  After a bit of algebra this gives the standard form of the soft factor, according to 
\eq{l19}{   { 1+z\zb\over \sqrt{2}}  \left[ \sum_{k=1}^m {Q_k^\text{out} \over z-z_k^\text{out}}- \sum_{k=1}^n {Q_k^\text{in} \over z-z_k^\text{in}} \right]  =    \sum_{k=1}^m  \left[{\omega Q^\text{out}_k p_k^\text{out}\cdot \eps^+\over p_k^\text{out}\cdot q} -  \sum_{k=1}^n {\omega Q^\text{in}_k p_k^\text{in}\cdot \eps^+\over p_k^\text{in}\cdot q} \right]~. }

\section{S-matrix in curved spacetime}\label{Sec:Paricle production in curved space}

The  AFS approach is well suited to studying the production of  particles in a time dependent background geometry.  We consider a free, real, massless  scalar field $\phi(x)$  propagating on a fixed metric $g_{\mu\nu}(x)$ that is taken to approach the Minkowski metric asymptotically in any direction.  The spacetime is also taken to be globally hyperbolic without black holes, although the latter could be incorporated with suitable modifications.   The computation that follows reproduces in an efficient way a result usually derived using the machinery of canonical quantization,  see e.g. \cite{Birrell:1982ix, DeWitt:1975ys}.  A detailed pedagogical treatment arriving at the same exponentiated formula we derive here may be found in \cite{Preskill}.

Some of our previous formulas need minor modification due to the fact that the background geometry, and hence Hamiltonian, is time dependent.  Equation \rf{r1} becomes 
\eq{r1a}{ \hat{S} = \lim_{\substack{t_i\rt -\infty\\t_f \rt +\infty}}e^{i\hat{H}_0 t_f}  T e^{-i\int_{t_i}^{t_f}  \hat{H}(t) dt  } e^{-i\hat{H}_0 t_i}~, }
with $\hat{H}_0 = \lim_{t\rt \pm \infty} \hat{H}(t)$.  Here $\hat{H}_0$ just corresponds to the Hamiltonian for a free field in Minkowski space with fixed time slices that approach $\scrI^\pm$ as the parameter $t$ tends to $\pm \infty$.     For massive particles, the analysis that follows goes through without change except that the early/late slices lie at fixed values of the ordinary Minkowski time coordinate $t$. 

In \cite{Adamo:2017nia,Adamo:2021rfq,Gonzo:2022tjm}, related methods are used to study the tree level S-matrix on classical backgrounds.

\subsection{Action and boundary conditions}

In this section it will be convenient to use the conventionally defined Klein-Gordon scalar product, 
\eq{p1}{ (\phi_1,\phi_2)_\text{KG}  & = - i\int_\Sigma d^3x \sqrt{h}  t^\mu ( \phi_1 \p_\mu \phi_2^* - \p_\mu \phi_1 \phi_2^*) \cr& = -2i(\phi_1, \phi_2^*)_\Sigma
}
where $t^\mu$ is the  future directed normal to the Cauchy surface $\Sigma$.   Setting the source to zero, the free scalar field  action is then 
\eq{p2}{ I & = I_\text{bulk} + I^G_\text{bndy}+ I^\text{free}_\text{bndy} \cr
& = {1\over 2}\int\! d^4x \sqrt{-g}  \phi_G \nabla^2 \phi_G  + {i\over 2} (\phib^f_-,\phib_+^*)^{\Sigma_f}_\text{KG}   -{i\over 2} ( \phib^i_+,\phib^*_-)^{\Sigma_i}_\text{KG}~.    }
The path integral over $\phi_G$ can be absorbed into the overall normalization of the S-matrix, which will later be fixed by unitarity.    We henceforth drop the overbars on $\phi$ since this is the only field that will appear, and we also suppress the KG subscript.  The action is thus written as 
\eq{p3}{ I = {i\over 2} (\phi_-^f,\phi_+^*)_{\Sigma_f}  -{i\over 2} ( \phi^i_+,\phi^*_-)_{\Sigma_i}~,  }
with boundary conditions 
\eq{p4}{ \phi_-(x) &= \phi_-^f(x) \quad {\rm on~} \scrI^+\cr
\phi_+(x) &= \phi_+^i(x) \quad {\rm on~} \scrI^-~.}
Our task is then to compute the inner products in \rf{p3} for a solution of the wave equation obeying the boundary conditions \rf{p4}.   As usual in problems of this type, it is useful to make reference to two distinct complete sets of modes solutions, related by Bogoliubov coefficients. 

\subsection{Bogoliubov relations}

For ease of notation it will be useful to work with a discrete set of basis functions; at the end of the computation we will convert back to the continuous plane wave basis. 

Let $\{u_i,u^*_i\}$ and $\{v_i,v^*_i\}$ be two complete sets of solutions of the wave equation, normalized with respect to the Klein-Gordon inner product as 
\eq{p5}{ & (u_i,u_j) = \delta_{ij}~,\quad (u^*_i,u^*_j)=-\delta_{ij}~,\quad (u_i,u^*_j)=0, \cr
& (v_i,v_j) = \delta_{ij}~,\quad (v^*_i,v^*_j)=-\delta_{ij}~,\quad (v_i,v^*_j)=0~.}
We will take  $u_i$ to be positive frequency on $\scrI^+$, and $v_i$ to be  positive frequency on $\scrI^-$.  The two sets are related as 
 \eq{p6}{ u_i &=\sum_j  (  \alpha_{ij}v_j+ \beta_{ij}v^*_j  ) \cr v_i&= \sum_j (\alpha^*_{ji} u_j-\beta_{ji} u^*_j  )~.  }
In an obvious matrix notation, the Bogoliubov coefficients obey the relations
 \eq{p7}{ \alpha \alpha^\dagger -\beta \beta^\dagger & =1, \cr
 \alpha^T \alpha^* -\beta^\dagger \beta  & =1, \cr
 \alpha \beta^T - \beta \alpha^T & = 0, \cr
 \alpha^T \beta^* -\beta^\dagger \alpha & = 0, }
 as obtained by substituting one line of \rf{p6} into the other and demanding consistency.   The solution $\phi(x)$ may be expanded in either set of modes,
\eq{p8}{ \phi  &= \sum_i \left( p_i u_i + p^\dagger_i u^*_i \right),  \cr
&= \sum_i \left( q_i v_i + q^\dagger_i v^*_i \right)~. }
The positive and negative frequency parts on $\scrI^\pm$ are then identified to be
\eq{p10}{ \phi^f_+ & = \sum_i  p_i u_i~,\quad 
\phi^f_-  = \sum_i  p^\dagger_i u^*_i \cr
\phi^i_+ & = \sum_i  q_i v_i~,\quad 
\phi^i_-  = \sum_i  q^\dagger_i v^*_i~.}
The boundary conditions \eqref{p4} then imply that the coefficients $\{p_i^\dag, q_i\}$ are the fixed data.

The problem to be solved is then the following: given $\{p_i^\dagger, q_i\}$, we need to compute $\{p_i,q_i^\dagger\}$.  This problem is straightforward to solve.  For example, we get one equation by writing $p_i =(\phi,u_i) $ and then expanding $\phi$ and $u_i$ in terms of $\{v_i,v^*_i\}$.  Doing this also for $q_i^\dagger$ we get a pair of equations,
 \eq{p8a}{ p &= \alpha^* q - \beta^* q^\dagger \cr q^\dagger &= \alpha^\dagger p^\dagger + \beta^T p }
 which are  easily solved with help from \rf{p7} to give  
 \eq{p9}{
 p& = {\alpha^T}^{-1} q - {\alpha^T}^{-1} \beta^\dagger p^\dagger\cr
 q^\dagger & = {\alpha }^{-1} p^\dagger +  {\alpha}^{-1} \beta  q~.
 }

\subsection{Evaluation of action}

The inner products in the action are straightforward to evaluate using the mode expansions of the previous section and \eqref{p9}. The two boundary terms are 
\eq{p11}{   \big(\phi^f_-,{\phi^f_+}^*\big) 
%&  =\sum_{ij} p_i^\dagger p_j (u_i^*,u_j^*) \cr
& = -p^\dagger p   \cr
& = - q \alpha^{-1} p^\dagger + p^\dagger \beta^* \alpha^{-1} p^\dagger \cr
\big(\phi^i_+,{\phi^i_-}^*\big) & = 
%\sum_{ij} q_i q_j^\dagger ( v_i,v_j) \cr
%& =
qq^\dagger \cr
&= q \alpha^{-1}p^\dagger + q \alpha^{-1} \beta q~.  }
This gives the on-shell action
\eq{p12 }{I &= {i\over 2} ( \phib_-^f,\phi_{+}^{f*})  -{i\over 2} ( \phib_+^i,\phi_{-}^{i*}) \cr
&= - i  \left(  q \alpha^{-1} p^\dagger   + {1\over 2} q \alpha^{-1} \beta  q - {1\over 2} p^\dagger \beta^* \alpha^{-1} p^\dagger  \right)~. }
The S-matrix operator is therefore found to be
\eq{p13}{ \hat{S} &= N :e^{i(I-I_0)} :\cr
& = N: e^{ \left(  q (\alpha^{-1}-\mathbb{1}) p^\dagger   + {1\over 2} q \alpha^{-1} \beta  q - {1\over 2} p^\dagger \beta^* \alpha^{-1} p^\dagger  \right) }:~, }
where we used that $I_0$ just subtracts off the result for $(\alpha= \mathbb{1},\beta=0). $   The normalization coefficient $N$ should be chosen such that $\hat{S}^\dagger \hat{S} = \mathbb{1}.$

We finally convert to the plane wave basis used in the rest of this paper.   The corresponding mode  solutions are written as
\eq{p14}{ u(\pv) &= {1\over \sqrt{2\omega_p} } e^{ip\cdot x} \quad {\rm on~} \scrI^+  \cr
v(\pv) &= {1\over \sqrt{2\omega_p} } e^{ip\cdot x} \quad {\rm on~} \scrI^- ~. }
These obey 
\eq{p15}{ \big( u(\pv),u(\pv') \big) & = - \big( u^*(\pv),u^*(\pv') \big) = (2\pi)^3 \delta^3(\pv-\pv') \cr
\big( v(\pv),v(\pv') \big) & = - \big( v^*(\pv),v^*(\pv') \big) = (2\pi)^3 \delta^3(\pv-\pv')~. }
To convert to the continuum case we use 
\eq{p16}{ \sum_i \rt \int\! {d^3p \over (2\pi)^3}~,   }
and the mode coefficients are related as 
\eq{p17}{ p_i^\dagger ~\rt ~ {b^\dagger (\pv) \over \sqrt{2\omega_p}}~,\quad  q_i ~\rt ~ {b(\pv) \over \sqrt{2\omega_p}} ~.}
We then have
\eq{p18}{ I- I_0  &=- i  \int\! {d^3p \over (2\pi)^3} {d^3p' \over (2\pi)^3}  {1\over \sqrt{2\omega_p}}  {1\over \sqrt{2\omega_{p'}}}  \Big[  \Big(  \alpha^{-1}(\pv,\pv')-(2\pi)^3 \delta^3(\pv-\pv') \Big) b(\pv) b^\dagger(\pv') \cr
&\quad\quad \quad\quad \quad\quad \quad\quad  +{1\over 2} (\alpha^{-1} \beta)(\pv,\pv') b(\pv) b(\pv')  -{1\over 2}  (\beta^* \alpha^{-1})(\pv,\pv') b^\dagger(\pv)b^\dagger(\pv')   \Big]~. }
The S-matrix is given as above by $\hat{S} = N: e^{i (I-I_0)}:$.  Unitarity fixes the value of $N$ up to a phase \cite{DeWitt:1975ys}, 
\eq{p19}{ N = {1\over \big( \det \alpha^\dagger \alpha\big)^{1/4}}~. }
Our result for the S-matrix agrees with \cite{Preskill}.  

It should be clear from our discussion in earlier sections that the extension of this result to include other types of fields and their interactions is straightforward in principle.  

\section{Discussion}
\label{discussion}

In this paper we have shown that the AFS approach to the S-matrix is  well suited for a variety of scattering problems, especially those  in which asymptotic symmetries are relevant. Here we only considered the simplest example of the latter, namely QED, but much work in recent years has developed the story of asymptotic symmetries and soft theorems in the context of Yang-Mills theory and gravity, e.g. \cite{Strominger:2013lka,He:2020ifr,He:2014laa}, as well as subleading soft theorems, e.g. \cite{Campiglia:2016hvg,Laddha:2017vfh}. We expect that the AFS approach is efficient in these contexts as well. Other closely related topics include handling IR divergences by implementing the Faddeev-Kulish approach, e.g.  \cite{Kapec:2017tkm,Himwich:2020rro,Nguyen:2023ibj,Choi:2017ylo,Nande:2017dba}, and the celestial holography program of expressing S-matrix amplitudes as correlators of a putative CFT on the celestial sphere; see \cite{Raclariu:2021zjz,Prema:2021sjp,Pasterski:2021raf,McLoughlin:2022ljp} for reviews.

\section*{Acknowledgements}
We thank Thomas Dumitrescu,   Temple He, Enrico Herrmann,   Andrea Puhm, Trevor Scheopner, and Andy Strominger for discussions. P.K. and R.M. are supported in part by the National Science Foundation grant PHY-2209700.

\appendix

\section{General boundary term analysis}
\label{GBT}

\subsection{The AFS generating functional in quantum mechanics}\label{Sec:QM Example}

Here we briefly review the main points of the AFS approach to scattering using a quantum mechanical example; few details change in field theory. The material here can be found largely in  Faddeev's chapter in \cite{Balian:1976vq}.

We consider a theory with Lagrangian
\eq{a2 1}{L = \frac{1}{2}\dot x^2 - \frac{1}{2}\omega^2x^2 - V(t, x)}
where we assume that the  potential turns off at early/late times. The canonical momentum is  $p = \dot x$ and we define the creation/annihilation modes by
\eq{a2 2}{x = \frac{1}{\sqrt{2\omega}}(a + a^\dag),\ \ \ \ p = -i\sqrt{\frac{\omega}{2}}(a - a^\dag).}
Note that this definition holds even in the interacting theory; the modes $a(t)$ and $a^\dag(t)$ will simply not evolve in time by only phases.

Given  these definitions we define the coherent states
\eq{a2 3}{\hat a|\alpha\rangle = \alpha|\alpha\rangle,\ \ \ \ \langle \alpha^\dag|\hat a^\dag = \langle\alpha^\dag|\alpha^\dag.}
These states, taking $\alpha$ and $\alpha^\dag$ to be independent, not necessarily related by conjugation, are normalized as 
\eq{a2 3-1}{
\langle\alpha^\dag|\alpha\rangle = e^{\alpha^\dag \alpha}.
}
This choice of normalization is such that the free vacuum has unit norm, $\langle 0|0\rangle = 1$, and that for any state $|\psi\rangle$ we have $\langle\psi|\hat a^\dag|\alpha\rangle = \frac{\p}{\p\alpha}\langle\psi|\alpha\rangle$ and $\langle \alpha^\dag|\hat a|\psi\rangle = \frac{\p}{\p\alpha^\dag}\langle\alpha^\dag|\psi\rangle$. Furthermore, the procedure for converting between coherent matrix elements and normal ordered operators follows immediately from this normalization. If $\hat{\mathcal{O}} = \sum_{n,m}\mathcal{O}_{n,m}(\hat a^{\dag})^n\hat a^m$ is any normal ordered operator, then
\eq{a2 3-2}{
\mathcal{O}(\alpha^\dag, \alpha) \equiv \langle \alpha^\dag|\hat{\mathcal{O}}|\alpha\rangle = e^{\alpha^\dag \alpha}\sum_{n,m}\mathcal{O}_{n,m}(\alpha^\dag)^n\alpha^m.
}
So to reconstruct the normal ordered operator $\hat{\mathcal{O}}$ from $\mathcal{O}(\alpha^\dag, \alpha)$, we need only replace $\alpha^\dag$ and $\alpha$ by $\hat a^\dag$ and $\hat a$, normal ordered, and cancel off the normalization \eqref{a2 3-1} of the coherent states.

Since $\hat H$ is not the free Hamiltonian, the time evolution of a coherent state will not remain coherent. We do, however, have
\eq{a2 4}{e^{-i\hat H_0t}|\alpha\rangle = |\alpha e^{-i\omega t}\rangle,\ \ \ \ \langle \alpha^\dag|e^{i\hat H_0 t} = \langle \alpha^\dag e^{i\omega t}|.}

Now, our goal is to compute the matrix elements of the S-matrix operator between coherent states:
\eq{a2 5}{S[\alpha^\dag, t_f; \alpha, t_i] &= \langle \alpha^\dag| e^{i\hat H_0t_f}Te^{-i\int_{t_i}^{t_f}\hat H\dr t}e^{-i\hat H_0 t_i}|\alpha\rangle\cr
&= \langle \alpha^\dag e^{i\omega t_f}| Te^{-i\int_{t_i}^{t_f}\hat H\dr t} |\alpha e^{-i\omega t_i}\rangle.}
Here we are considering a finite time transition, but will have in mind taking $t_f \rightarrow \infty$ and $t_i \rightarrow -\infty$ at the end. We again stress that the states $|\alpha\rangle$ and $\langle \alpha^\dag|$ are independent, so $\alpha$ and $\alpha^\dag$ need not be related by conjugation.

The latter form of \eqref{a2 5} is important for us because it allows us to compute the transition amplitude using a path integral with modified boundary conditions, rather than some complicated operator insertions:
\eq{a2 6}{
S[\alpha^\dag, t_f; \alpha, t_i] = \int_{a(t_i) = \alpha e^{-i\omega t_i}}^{a^\dag(t_f) = \alpha^\dag e^{i\omega t_f}}[\mathcal{D}x\mathcal{D}p] e^{iI[x,p]}
}
where $I[x, p]$ should be the phase space action 
\eq{a2 7}{
I[x, p] &= -ia^\dag a|_{t_f} + \int_{t_i}^{t_f}(ia^\dag \dot a - H)\dr t\cr
&= -\frac{i}{2}(a^\dag a|_{t_f} + a^\dag a|_{t_i}) + \int_{t_i}^{t_f}\left(\frac{1}{2}(p\dot x - x\dot p) - H\right)\dr t.
}
This is the unique choice of kinetic and boundary terms such that $I$ is compatible with the scattering boundary conditions in \eqref{a2 6} and also\footnote{This latter condition excludes the possibility of adding to $I$ a functional of only the boundary conditions, which would not hamper the variational principle.}
\eq{a2 8}{
    \frac{\p}{\p \alpha}iI_\text{shell} = a^\dag(t_i)e^{-i\omega t_i},\ \ \ \ \frac{\p}{\p \alpha^\dag}iI_\text{shell} = a(t_f)e^{i\omega t_f}
}
where $I_\text{shell}$ is the on-shell action. This is also the boundary term obtained by K\"ahler quantization of the harmonic oscillator, and by the explicit slicing construction of the path integral.

The conditions to this point imply that
\eq{a2 9}{
    \frac{\p^m}{\p\alpha^{\dag\,m}}\frac{\p^n}{\p \alpha^n}S[\alpha^\dag, t_f; \alpha, t_i] &= \langle \alpha^\dag e^{i\omega t_f}| (\hat a e^{i\omega t_f})^m Te^{-i\int_{t_i}^{t_f}\hat H\dr t} (\hat a^\dag e^{-i\omega t_i})^n |\alpha e^{-i\omega t_i}\rangle \cr
    &= \langle \alpha^\dag| \hat a^m \left( e^{i\hat H_0 t_f} Te^{-i\int_{t_i}^{t_f}\hat H\dr t} e^{-i\hat H_0 t_i} \right) \hat a^{\dag\, n}|\alpha\rangle.
}
Hence, using that $|\alpha = 0\rangle$ is the free vacuum, we see that the coherent state matrix elements $S[\alpha^\dag, t_f; \alpha, t_i]$ form the generating functional of standard ``point particle'' S-matrix elements. Indeed, in \cite{Schwinger:1953zza} Schwinger obtained the S-matrix of electrodynamics by first constructing the coherent state matrix elements. When generalized to field theory, the demonstration \eqref{a2 9} can be viewed as an operator-based argument for the equivalence of the AFS and LSZ prescriptions.

\subsection{Boundary term cancellation and efficient computation}\label{Sec:Boundary Term Cancellation}

In the main text, the boundary terms, which we understood systematically in the previous section, were ``evaluated'' by explicitly rewriting them as bulk terms\footnote{With the exception of the free boundary term, which yields the forward scattering portion of the S-matrix, as we will see.  }. The details of this computation may seem special to the types of fields and theories that we have considered in this paper, but the mechanism is completely generic.

In this appendix we describe how this mechanism occurs, using the quantum mechanical example of the previous section to simplify expressions since few details change in field theory. Understanding the general mechanism also yields a very efficient approach to dealing with the boundary terms in perturbation theory, whereby one guarantees they are cancelled without needing to write them down explicitly.

Starting with the action \eqref{a2 7} as it appears in the path integral \eqref{a2 6}, we first shift the integration variables to $x = x_{cl} + \tilde x$ and $p = p_{cl} + \tilde p$ where $x_{cl}, p_{cl}$ are the solution to classical equations of motion satisfying the scattering boundary conditions present in \eqref{a2 6}. Hence $\tilde x$ and $\tilde p$ obey vanishing scattering boundary conditions: $\tilde a(t_i) = 0$ and $\tilde a^\dag(t_f) = 0$.

Expanding the action about the classical solution,
\eq{a1 1-1}{
I[x, p] = I[x_{cl}, p_{cl}] + \int\dr t\left[ \frac{\delta I}{\delta x(t)}|_{x_{cl},p_{cl}}\tilde x(t) + \frac{\delta I}{\delta p(t)}|_{x_{cl},p_{cl}}\tilde p(t) \right] + \mathcal{O}(\tilde x^2, \tilde x \tilde p, \tilde p^2). 
}
The $I$ here includes the boundary terms in \eqref{a2 7}, which have the important property that they are {\em linear} in the perturbations $\tilde x, \tilde p$. Explicitly,
\eq{a1 1-2}{
I_\text{bndy}[x, p] = -\frac{i}{2}(a^\dag_{cl} a_{cl}|_{t_f} + a^\dag_{cl} a_{cl}|_{t_i}) - \frac{i}{2}(a^\dag_{cl}\tilde a|_{t_f} + \tilde a^\dag a_{cl}|_{t_i}).
}
The second set of terms, linear in perturbations, were designed from the beginning to make the linear terms in \eqref{a1 1-1} vanish; this was the demand of a good variational principle.

Hence we find that\footnote{In the path integral, we could have in principle shifted by any configuration which obeys the boundary conditions, not just a classical solution. Here we see that if we had chosen any other configuration, the fluctuating fields $\tilde x, \tilde p$ would acquire tadpoles in the loop computation. Removing the tadpoles shifts us back over to a classical solution.}
\eq{a1 1-3}{
I[x, p] = I[x_{cl}, p_{cl}] + \mathcal{O}(\tilde x^2, \tilde x\tilde p, \tilde p^2)
}
and the only boundary terms remaining are the ones evaluated on the classical solution, i.e. the first pair of terms in \eqref{a1 1-2}. While we have argued here on general principles that the boundary term cancellation must occur, it's also simple to see this by direct computation in the present quantum mechanical example.

If the action splits into free and interacting terms, as \eqref{a2 1} does, we may also use this same approach to efficiently compute the on-shell action in the first term of \eqref{a1 1-3}. Write $x_{cl} = \overline x + x_G$ and $p_{cl} = \overline p + p_G$ where $\overline x, \overline p$ are free solutions saturating the boundary conditions, meaning $a_G(t_i) = 0$ and $a^\dag_G(t_f) = 0$. Then
\eq{a1 1-4}{
I[x_{cl}, p_{cl}] = I_\text{int}[x_{cl}, p_{cl}] + \overline I[\overline x, \overline p] + \int\dr t\left[ \frac{\delta \overline I}{\delta x(t)}|_{\overline x,\overline p} x_G(t) + \frac{\delta \overline I}{\delta p(t)}|_{\overline x, \overline p}p_G(t) \right] + \mathcal{O}(x_G^2, x_G p_G, p_G^2).
}
As before, the boundary terms are only linear in the perturbation around $\overline x, \overline p$, and so completely cancel against the terms in the bulk linear in the perturbations.

Considering specifically \eqref{a2 1}, to evaluate the on-shell action we write (using $p_{cl} = \dot x_{cl}$)
\eq{a1 1-5}{
I[x_{cl}] &= -\int V(t, x_{cl})\dr t + I_\text{bndy} + \int\left[ -\frac{1}{2}x_{cl}\ddot x_{cl} - \frac{1}{2}\omega^2 x^2_{cl} \right]\\
&= -\int V(t, x_{cl})\dr t + I_{\text{bndy}} - \int \left[ \frac{1}{2}(x_G \ddot {\overline x} + \overline x \ddot x_G) + \omega^2 x_G \overline x \right]\dr t - \frac{1}{2}\int x_G\left[ \ddot x_G + \omega^2 x_G \right]\dr t.
}
To use the free equations of motion to cancel the middle term, it's necessary to integrate the derivatives from $x_G$ onto $\overline x$. This is what produces the boundary terms that cancel the linear-in-perturbation terms of $I_\text{bndy}$, which we argued must happen on general principles, but which is also easily checked explicitly in this example.

Practically, because this cancellation must occur, we can just drop the terms linear in $x_G$ to find
\eq{a1 1-6}{
I[x_{cl}] = -i\alpha^\dag \alpha + \int \left[ \frac{1}{2}x_G \p_x V(t, x_{cl}) -  V(t, x_{cl}) \right]\dr t.
}
Here the remaining, free, boundary terms are completely determined by the boundary conditions, and we have used the equations of motion to write $\ddot x_G + \omega^2 x_G = -\p_x V'(t, x_{cl})$. Computing the S-matrix in normal ordered form via \eqref{a2 3-2}, we see that dividing out by the coherent state norm just serves to cancel the free boundary term.

\section{Generating functional approach to the soft theorem}\label{Sec:Sourcing Approach}

In section \ref{LGT and soft} we showed how the soft theorem could be obtained by considering the commutator of the operator form of the S-matrix with the so-called soft photon creation operator $\hat N_z = \p_z \hat N$. Here we show how the same result can be obtained without first converting to the normal ordered form.

Before getting to the details, we can see the basic mechanism by which the soft theorem will arise in the generating functional approach. Taking a $\Phi_0$ derivative of the S-matrix will produce an insertion of its canonical conjugate, roughly $\hat N$, as a matter of definition. On the other hand, we could use \eqref{l13}, which relates $\Phi_0$ derivatives to hard insertions. Conversely, any derivatives with respect to the hard particle sources are accompanied by an $e^{iQ\Phi_0(z, \overline z)}$ factor on which the $\Phi_0$ derivatives can act.

To get the details correct, we need to be careful about what operators our derivatives are inserting. As we will see, $\Phi_0$ does not source exactly the right hand side of the soft theorem \eqref{l18}, so we first need to find the canonical conjugate to $\p_zN$ and write $\Phi_0$ in terms of it. For this purpose, we will require the portion of the symplectic form involving $\Phi_0$:
\eq{ap 5 1}{
\Omega_{\scrI^+} = \int\dr^2 z\delta(2\p_z\p_{\overline z}\Phi_0)\wedge\delta N.
}
This is only the soft sector's contribution to the symplectic form living on $\scrI^+$, but it will be all we need since $N$ and $\Phi_0$ commute with all bulk fields \cite{He:2014cra}. This symplectic form tells us that a $\Phi_0$ derivative of the S-matrix \eqref{l13} will produce an insertion\footnote{Strictly, since $\Phi_0$ is forced to be antipodally matched at early and late times, this derivative produces an antipodally matched pair of insertions, which will later lead to the commutator in \eqref{l18} instead of a single operator insertion.} of $i(-2\p_z\p_{\overline z}\hat N) = -i2\p_z\p_{\overline z}\hat N$, the extra $i$ being the same imaginary unit multiplying the action in the path integral -- we can ground ourselves by noting that in quantum mechanics, the symplectic form $\Omega = \delta p \wedge \delta x$ leads to $\frac{\p}{\p p}\langle p|\psi\rangle = \langle p|(-i\hat x)|\psi\rangle$.

From this symplectic form we see that $\pi_N = 2\p_z\p_{\overline z}\Phi_0$ is the conjugate to $N$. Inverting this relation\footnote{We use that shifting $\Phi_0$ by a constant on the celestial sphere is a null direction of the symplectic form, and hence is non-physical, to fix the possibility of an additive constant in this inversion.},
\eq{ap 5 2}{
\Phi_0 = \frac{1}{4\pi}\int\dr^2 w \pi_N(w, \overline w)\ln|z - w|^2.
}
So if we used this relation to replace $\Phi_0$ in favor of $\pi_N$ in the S-matrix, we could take $\pi_N$ derivatives to produce insertions of $-i\hat N$.

To generate insertions of $\p_z \hat N$, we write \eqref{ap 5 1} as
\eq{ap 5 4}{
\Omega_{\scrI^+} = \int\dr^2 z\delta(-2\p_{\overline z}\Phi_0) \wedge \delta \p_z N,
}
so the conjugate to $\p_z N$ is $\pi_{N_z} = -2\p_{\overline z}\Phi_0$. But this also means that $\pi_N = -\p_z\pi_{N_z}$ and hence
\eq{ap 5 7}{
\Phi_0 = \frac{1}{4\pi}\int\dr^2 w \frac{\pi_{N_z}(w, \overline w)}{w - z}.
}

Now, since $\Phi_0$ is antipodally matched, we see that
\eq{ap 5 8}{
i\frac{\delta}{\delta \pi_{N_z}(z, \overline z)}&S[b, b^\dag, c, c^\dag, \tilde{\overline A}, \Phi_0[\pi_{N_z}]]\cr
&= \langle b^\dag, c^\dag, \tilde{\overline A}_-|\p_z\hat N\hat S - \hat S\p_z\hat N|b, c, \tilde{\overline A}_+\rangle = \langle b^\dag, c^\dag, \tilde{\overline A}_-|[\p_z \hat N, \hat S]|b, c, \tilde{\overline A}_+\rangle.
}

While the content of the soft theorem in the generating function approach is contained in \eqref{l13}, to obtain specifically the standard formula \eqref{l18}, it's simplest to consider a specific process.
On the one hand, by definition we have
\eq{ap 5 8-1}{
\mathcal{A} = i\frac{\delta}{\delta\pi_{N_z}(z, \overline z)}\prod^N_{k \text{ in}}&\left( {2\omega_{p_k}(2\pi)^3}\frac{\delta}{\delta b_k(\vec p_k)} \right)\prod^M_{\ell \text{ out}}\left( {2\omega_{q_\ell}}{(2\pi)^3}\frac{\delta}{\delta b^\dag_\ell(\vec q_\ell)} \right)S|_0 \cr
&= \langle q_1,\ldots,q_M|[\p_z \hat N(z, \overline z), \hat S]|p_1,\ldots,p_N\rangle.
}
On the other hand, using \eqref{l13} and doing the mode derivatives first, we find
\eq{ap 5 8-2}{
\mathcal{A} &= i\frac{\delta}{\delta \pi_{N_z}(z, \overline z)}\langle q_1,\ldots, q_M|\hat S|p_1,\ldots,p_N\rangle\exp\left[-i\sum_{k\text{ in}}^NQ_k\Phi_0(z_k, \overline z_k) + i\sum_{\ell\text{ out}}^MQ_\ell \Phi_0(z_\ell, \overline z_\ell) \right]\Bigg|_{\Phi_0 = 0}\cr
&= \frac{1}{4\pi}\left[ \sum_{k\text{ in}}^N\frac{Q_k}{z - z_k} - \sum_{\ell\text{ out}}^M\frac{Q_\ell}{z - z_\ell} \right]\langle q_1,\ldots, q_M|\hat S|p_1,\ldots,p_N\rangle.
}
Equality of these two computations establishes the soft theorem in the form \eqref{l18}.

\section{ Equivalence of the  GKP/W and BDHM prescriptions for AdS  boundary correlators   }
\label{sec:AdSCFT}

As noted in the introduction, there are two prescriptions for computing boundary correlators in asymptotically  AdS  spacetimes.  The GKP/W prescription \cite{Gubser:1998bc,Witten:1998qj}, in which one views the path integral with specified boundary conditions as a generating functional of boundary correlators, and the BDHM \cite{Banks:1998dd} prescription in which boundary correlators are defined as the limit of  bulk correlators as the operator locations are taken to the boundary.   As we show in this appendix, their equivalence can be established in a manner parallel to how we demonstrated equivalence of the LSZ and AFS prescriptions for the Minkowski space S-matrix.

For definiteness we consider a real scalar field in Euclidean signature,
\eq{ap1}{
	I[\phi] = \int_M  \! d^{d+1} x \sqrt{g} \, \left[  \frac{1}{2}  \nabla^\mu \phi\nabla_\mu \phi +\frac{1}{2} m^2\phi^2 +V(\phi) -J\phi \right]+I_\text{bndy}~.  }
We write the coordinates as $x^\mu=(z,y^i)$ where $z$ is the radial coordinate.  
The metric is asymptotically AdS$_{d+1}$ and  takes the form
\eq{ap2}{ ds^2 = {dz^2 \over z^2} +  h_{ij}(z,y) dy^i dy^j ~,
 }
with 
\eq{ap2a}{ h_{ij}(z,y) =  {1\over z^2 }g^{(0)}_{ij}(y) dy^i dy^j + \ldots~  }
where the boundary is at $z=0$, and $\ldots$ denote terms higher order in $z$.  We proceed by imposing a cutoff at $z=\eps$, which is eventually taken to zero, and impose boundary conditions on the scalar field as 
\eq{ap3}{ \phi(z,y)\big|_{z=\eps} = \phib(y) \eps^{d-\Delta}~,}
where $\Delta $  is the larger root of $m^2 = \Delta(\Delta-d)$.   Given this Dirichlet boundary condition, $I_\text{bndy}$ in \rf{ap1} is taken to be some local expression built out of $\phib(y)$, and is partly determined by the requirement that it cancel the local divergences that arise in the $\eps \rt 0$ limit.   The fact that $I_\text{bndy}$ is local implies that it does not affect boundary correlation functions at noncoincident points, and hence can be ignored for what follows.

Following our S-matrix discussion we focus on the path integral in the presence of sources and nontrivial boundary conditions,
\eq{ap4}{ Z[\phib,J] = \int\! {\cal D} \phi \, e^{-I[\phi]}~.}
We now define $\phib_J$ as the solution of the sourced free field equation that obeys our boundary condition,
\eq{ap5}{ ( -\nabla^2+m^2) \phib_J(x)  = J(x)~,\quad \phib_J(z,y)\big|_{z=\eps} = \phib(y) \eps^{d-\Delta}~. }
To solve this we write $\phib_J$ in the form 
\eq{ap6}{ \phib_J(x) = \phib(x) + \int \!d^{d+1}x' \sqrt{g} \, G(x;x') J(x')~, }
where  $\phib(x)$  is defined as the solution of 
\eq{ap8}{ (-\nabla^2+m^2) \phib(x) = 0 }
with boundary condition 
\eq{ap9}{ \phib(z,y)\big|_{z=\eps} = \phib(y) \eps^{d-\Delta}~. }
Here $G(x,x')$ is the AdS bulk-bulk propagator obeying 
\eq{ap7}{ (-\nabla^2 +m^2)G(x,x') = {1\over \sqrt{g} } \delta^{(d+1)}(x-x')}
and which vanishes at the boundary. 
We further express $\phib(x) $ using the bulk-boundary propagator $K(x;y')$ as
\eq{ap10}{ \phib(x) = \int_{\p \text{AdS}} \! d^dy' \sqrt{h} K(x;y') \phib(y')~, }
where  $K$ is defined via
\eq{ap11}{ (-\nabla^2+m^2)K(x;y')=0~,\quad  K(x;y')\big|_{z=\eps} = {\eps^{d-\Delta }\over \sqrt{h} } \delta^{(d)}(y-y')~. }
The general field configuration contributing to the path integral now takes the form 
\eq{ap12}{ \phi(x) = \phib_J(x) + \phi_G(x) }
where $\phi_G(x)\big|_{z=\eps}=0$.
After some algebra and using the stated properties of $\phib_J$ and $\phi_G$ the action takes the form 
\eq{ap13}{ I &= \int_\text{AdS}\! d^{d+1}x \sqrt{g} \Big[ {1\over 2} \nabla^\mu \phib_J \nabla_\mu \phib_J -{1\over 2} m^2 \phib_J^2 + \phib_J \nabla^2 \phib_J  \cr
&\quad\quad\quad\quad\quad\quad\quad  +{1\over 2} \nabla^\mu \phi_G \nabla_\mu \phi_G +{1\over 2} m^2 \phi_G^2+ V(\phib_J+\phi_G) \Big]~.  }
We distinguish the contributions in the two lines by writing  $I =I_0[\phib_J] + I_G[\phib_J,\phi_G] $.
Now, $I_0[\phib_J] $ is independent of  $\phi_G$ and so can be pulled out of the path integral,
\eq{ap14}{ Z[\phib,J] = e^{-I_0[\phib_J]}  \int\! {\cal D}\phi_G e^{-I_G[\phib_J,\phi_G] }~.}
We note the two key properties of  $Z_G[\phib_J] = \int\! {\cal D}\phi_G e^{-I_G[\phib_J,\phi_G]} $.   First, it depends on $\phib$ and $J$ solely through $\overline\phi_J$.  Second, order by order in perturbation theory, $Z_G[\phib_J]$ is a functional of $\phib_J$ without any derivatives acting on $\phib_J$. 

We are now ready to compare the two forms of boundary correlators. In the BDHM prescription we set $\phib(y)=0$, in which case $Z[0,J]$ becomes the generating functional for bulk correlators,
\eq{ap15}{ Z[0,J]= \sum_{n=1}^\infty {(-1)^n\over n!} \left[ \prod_{i=1}^n \int\! d^{d+1}x_i \sqrt{g(x_i)} \right] G_n(x_1,\ldots ,x_n) J(x_1)\ldots J(x_n)~.  }
The boundary correlators are then extracted as  
\eq{ap16}{ \langle {\cal O}(y_1) \ldots {\cal O}(y_n)\rangle = (2\Delta-d)^n\lim_{z_i \rt \epsilon} z_1^{-\Delta } \ldots z_n^{-\Delta} G_n(x_1,\ldots ,x_n)~. }
Now, the diagrams contributing to $G_n(x_1,\ldots , x_n)$ all have a bulk-bulk propagator emanating from each bulk point $x_i$.   Furthermore, the bulk-bulk propagator is related to the bulk-boundary propagator as 
\eq{ap17}{ K(x;y') = (2\Delta-d)\lim_{z'\rt \epsilon} z'^{-\Delta} G(x,x')~. }
The factors of $2\Delta -d$ are discussed in \cite{Klebanov:1999tb} among other places. 
From this we see that if in \rf{ap15} we make the replacement 
\eq{ap18}{ \int \! d^{d+1} x_i \sqrt{g(x_i)} G(x'_i,x_i)J(x_i)  ~\rt ~ \int \! d^{d} y_i \sqrt{h(y_i)}K(x'_i;y_i) \phib(y_i)   }
then $Z[0,J]$ turns into the generating functional of boundary correlators, 
\eq{ap19}{Z[0,J]\big|_{GJ \rt K\phib} =   \sum_{n=1}^\infty {(-1)^n\over n!} \left[ \prod_{i=1}^n \int\! d^{d}y_i \sqrt{h(x_i)} \right] \langle {\cal O}(y_1) \ldots {\cal O}(y_n)\rangle_\text{BDHM} \phib(y_1)\ldots \phib(y_n)~.  }
On the other hand, in the GKP/W prescription we set $J=0$, compute $Z[\phib,0]$, and identify it as the generating functional of boundary correlators, 
\eq{ap20}{Z[\phib,0] =   \sum_{n=1}^\infty {(-1)^n\over n!} \left[ \prod_{i=1}^n \int\! d^{d}y_i \sqrt{h(y_i)} \right] \langle {\cal O}(y_1) \ldots {\cal O}(y_n)\rangle_\text{GKP/W} \phib(y_1)\ldots \phib(y_n)~.  }
To establish that $Z[0,J]\big|_{GJ \rt K\phib}=Z[\phib,0]$, and hence that 
\eq{ap20a}{ \langle {\cal O}(y_1) \ldots {\cal O}(y_n)\rangle_\text{BDHM}= \langle {\cal O}(y_1) \ldots {\cal O}(y_n)\rangle_\text{GKP/W}~, }
we observe that $\phib_J$ is the same whether we consider $(\phib,J=0)$ or $(\phib=0,J)\big|_{GJ \rt K\phib}$.  This fact, together with the no derivative property of $Z_G[\phib_J]$ is sufficient to establish the equivalence of the two prescriptions for computing boundary correlators.   The same argument may be carried out for theories with more general matter content and other boundary conditions.

\section{Classical scattering solutions}
\label{classical}

At tree level, the S-matrix is obtained by computing the classical action evaluated on a classical solution of the equations of motion.  However, the classical solution of interest is not one that would be considered in a standard classical physics problem, because the fields are subject to ``in-out" boundary conditions. For instance, for a real scalar field the classical solution that enters into the S-matrix is generically complex.   Furthermore, the large $r$ asymptotics are nonstandard.

To illustrate this we consider computing the gauge field sourced by a charged particle moving along a prescribed timelike trajectory $X^\mu(\tau)$.   Assuming Lorenz gauge, we need to solve
\eq{A1}{ \nabla^2 A^\mu(x) = J^\mu(x) }
for the appropriate conserved current,
\eq{A2}{ J^\mu(x) =\int\! d\tau {dX^\mu(\tau) \over d\tau}\delta^4(x-X(\tau))~. }
If we integrate \rf{A1} using the retarded propagator we arrive at the standard Li\'enard-Wiechert potentials, which have the property that the Cartesian components  $A_\mu$ have $1/r$ falloff near $\scrI^\pm.$   The same falloff would arise from use of the advanced propagator.  Such a $1/r$ falloff is part of the definition of the usual phase space for the electromagnetic field\footnote{Note that this assumes the absence of massless charged particles.  In the presence of massless charged particles that enter and exit through $\scrI^\pm$ the falloff is weaker; see e.g. \cite{Prabhu:2022zcr}.}.

However, in the S-matrix context we should instead use the Feynman propagator, which unlike the retarded and advanced propagators, has support off of the light cone, and is also complex valued.  Defining  (see \cite{Bjorken:1965zz}) 
\eq{A3}{  G(x) & = {1\over 4\pi |\xv|} \big[ \delta(|\xv|-t) - \delta(|\xv|+t)\big] \cr
G_1(x) & = {1\over 4\pi^2 |\xv|  } \left( P {1\over |\xv|-t} + P {1\over |\xv|+t} \right)  }
the retarded and advanced propagators are
\eq{A4}{G_\text{ret}(x) & = G(x)\Theta(t) \cr
G_\text{adv}(x) & = -G(x)\Theta(-t)   }
while the Feynman propagator is 
\eq{A5}{ G_F(x) & =  -{i\over 2} G_1(x)  +{1\over 2} {\rm sgn}(t) G(x) }
These obey $-\nabla^2 G_\text{ret}(x) =-\nabla^2 G_\text{adv}(x) =-\nabla^2 G_F(x) =\delta^{(4)}(x).  $

As a simple illustrative example (see \cite{AtulBhatkar:2021txo} for a more systematic study), consider a particle which is at rest at the origin for $t<0$, and then moves at constant velocity $\vec{v} =v \hat{z}$ for $t>0$.  The gauge field sourced by this current gets contributions from both the $G_1$ and $G$ terms in \rf{A5}; since the latter just gives the usual retarded plus advanced Li\'enard-Wiechert potential we focus on the $G_1$ contribution.  A straightforward computation gives
\eq{A6}{ A_{G_1}(x) & = {i\over 8\pi^2}   { 1\over |\xv| }  \ln\left|  {t-|\xv| \over t+|\xv|}\right| dt \cr
& -{i\over 8\pi^2} {1\over \sqrt{x^2+y^2+\gamma^2(z-vt)^2}}  \ln\left| {\gamma(t-vz)- \sqrt{x^2+y^2+\gamma^2(z-vt)^2}  \over  \gamma(t-vz)+ \sqrt{x^2+y^2+\gamma^2(z-vt)^2}  }  \right| \gamma( dt-vdz )~, \cr    }
with the top and bottom lines coming from the $t<0$ and $t>0$ parts of the current. Notice that if we set $v=0$ the two contributions cancel, and the full gauge field would just be the usual stationary Coulomb field.  To study the asymptotics near $\scrI^+$ we write  
\eq{A&}{ x^2+y^2 &= r^2 \sin^2\theta \cr
z& = r \cos\theta \cr
t & =u+r }
and  take $r\rt \infty $ at fixed $u$.  This gives the leading behavior
\eq{A7}{  A_{G_1}(x) &  \approx   {i\over 8\pi^2}   { 1\over r }  \ln\left|  { u \over 2r }  \right| (dr+du) \cr
 & -{i\over 8\pi^2}  {1\over v\cdot q}{1\over r} \ln \left| { u \over 2 (v\cdot q)^2 r}\right| \big(  v\cdot q dr +\gamma(du+ rv \sin \theta d\theta) \big) }
where $v\cdot q = \gamma(1-v\cos \theta)$.  We note in particular that this implies that the angular component behaves as $A_\theta \sim \ln r$ rather than $r^0$ as is usually assumed.   

This asymptotic behavior is presumably representative of the generic solution involving classical solutions obeying the S-matrix boundary conditions.  As we have noted,  the nonstandard asymptotic behavior of these classical solutions is a reflection of the fact that  they correspond to in-out matrix elements of the field operator, as opposed to expectation values.   It does however imply that we need to exercise some care in taking the large $r$ limit, by taking the limit only after computing relevant quantities.

\bibliographystyle{bibstyle2017}
\bibliography{collection}

\end{document}